\documentclass[a4paper,fleqn,usenatbib]{mnras}

\usepackage{newtxtext,newtxmath}

\usepackage[T1]{fontenc}
\usepackage{ae,aecompl}

\usepackage[version=3]{mhchem}
\usepackage{placeins}
\usepackage{enumitem}
\usepackage{rotating}
\usepackage{graphicx}	
\usepackage{subcaption}
\usepackage{amsmath}	
\usepackage{longtable}
\usepackage{multirow}

\graphicspath{ {C:/Users/Student/Documents/1. PhD Work/2. Analysis of Hollands Data/Write Up} }
\title[Modelling the observed pollutant population]{Bayesian constraints on the origin and geology of exo-planetary material using a population of externally polluted white dwarfs}

\author[J. H. D. Harrison et al.]{
John H. D. Harrison$^1$ \thanks{E-mail: jhdh2@cam.ac.uk}, Amy Bonsor$^1$\thanks{E-mail: abonsor@cam.ac.uk}, Mihkel Kama$^{1,2,3}$, Andrew M. Buchan$^1$, \newauthor
Simon Blouin$^4$ and Detlev Koester$^5$
\\
$^1$Institute of Astronomy, University of Cambridge, Madingley Road, Cambridge, CB3 0HA, UK\\
$^{2}$ Department of Physics and Astronomy, University College London, Gower Street, London, WC1E 6BT, UK\\
$^{3}$ Tartu Observatory, University of Tartu, Observatooriumi 1, 61602 T\~{o}ravere, Tartumaa, Estonia\\
$^4$Los Alamos National Laboratory, PO Box 1663, Los Alamos, NM 87545, USA \\
$^5$Institut f\"{u}r Theoretische Physik und Astrophysik, University of Kiel, 24098 Kiel, Germany
}

\date{Accepted XXX. Received YYY; in original form ZZZ}

\pubyear{2020}

\begin{document}
\label{firstpage}
\pagerange{\pageref{firstpage}--\pageref{lastpage}}
\maketitle

\begin{abstract}
White dwarfs that have accreted planetary bodies are a powerful probe of the bulk composition of exoplanetary material. In this paper, we present a Bayesian model to explain the abundances observed in the atmospheres of 202 DZ white dwarfs by considering the heating, geochemical differentiation, and collisional processes experienced by the planetary bodies accreted, as well as gravitational sinking. The majority (>60\%) of systems are consistent with the accretion of primitive material. We attribute the small spread in refractory abundances observed to a similar spread in the initial planet-forming material, as seen in the compositions of nearby stars. A range in Na abundances in the pollutant material is attributed to a range in formation temperatures from below 1,000\,K to higher than 1,400\,K, suggesting that pollutant material arrives in white dwarf atmospheres from a variety of radial locations. We also find that Solar System-like differentiation is common place in exo-planetary systems. Extreme siderophile (Fe, Ni or Cr) abundances in 8 systems require the accretion of a core-rich fragment of a larger differentiated body to at least a 3$\sigma$ significance, whilst one system shows evidence that it accreted a crust-rich fragment. In systems where the abundances suggest that accretion has finished (13/202), the total mass accreted can be calculated. The 13 systems are estimated to have accreted masses ranging from the mass of the Moon to half that of Vesta. Our analysis suggests that accretion continues for 11Myrs on average.
\end{abstract}





\section{Introduction}

Knowledge of the bulk composition of exo-planetary material is crucial to understanding the interiors of rocky exoplanets, as well as their geological processes, and their potential to be habitable. White dwarfs that have accreted planetary bodies provide a powerful means to investigate the bulk composition of exo-planetary material. Metal lines in the spectra of these polluted white dwarfs can be used to infer the chemical abundances of the accreted material. In this paper, we study a large sample of such objects to constrain the geological origin and mass of the polluting exo-planetesimals.

Externally polluted white dwarfs are cool white dwarf stars with metal features in their spectra \citep{JuraYoung2014}. Strong metal absorption lines have been detected in more than one thousand cool white dwarfs \citep{Weidemann1960, Zuckerman1998, Kepler2016, Hollands2017, Coutu2019}. These metals sink out of sight on timescales of days to millions of years  \citep{Fontaine1979, Koester09}.

The cooling ages of these white dwarfs are considerably longer than these timescales (of the order tens of millions to billions of years), therefore, these metals must have been accreted onto the white dwarfs relatively recently.

Planetary material is thought to arrive in the white dwarf's atmosphere from a remnant planetary system that has survived the star's post-main sequence evolution as a natural by product of the evolution of the system \citep{Veras_review}. Dynamical instabilities and planetesimal scattering caused by any surviving planets during the white dwarf phase can lead to planetary bodies being perturbed onto star-grazing orbits \citep{DebesSigurdsson, bonsor11, debesasteroidbelt,Mustill2018}, where they are likely to become tidally disrupted, form an accretion disc, and be accreted onto the white dwarf \citep{Jurasmallasteroid, Veras_tidaldisruption1, Veras_tidaldisrupt2}. Wide binary companions \citep{bonsor_wdbinary, Hamers2016, Petrovich2017, Stephan2017}, the liberation of exo-moons \citep{Payne2017}, and the ohmic heating of asteroids \citep{Bromley2019} have also been suggested as potential explanations for the pollution of white dwarfs. Although the exact mechanism by which material is transported to the white dwarf is not established, observations of disintegrating planetesimals orbiting polluted white dwarfs \citep{Vanderburg2015}, a giant planet orbiting a white dwarf \citep{Vanderburg2020} and observations of discs around polluted white dwarfs \citep{Farihi_review} have shown that the prevailing explanation for the presence of metal lines in the spectra of many white dwarfs is due to the accretion of rocky exo-planetary material \citep{Jurasmallasteroid, JuraYoung2014}.

The exact nature of the discs around white dwarfs is not yet understood \citep{Farihi_review}. The accretion event lifetimes, the amount of time accretion occurs for during a given event, have been estimated to be of the order of $10^{4}$ to $10^{6}$ years by \cite{Girven2012}. However, \cite{Wyatt2014} estimated that the disc lifetimes were potentially as short as 20 years due to variation between the accretion rates observed for cool DZ white dwarfs and warm DAZ white dwarfs.  Theoretical considerations of Poynting-Robertson drag driven accretion suggest that disc lifetimes for discs with a mass of $\sim$$10^{19}$\,kg are of the order of millions of years \citep{rafikov2, Rafikov1}. Constraining the lifetimes of white dwarf accretion events more precisely is vital if we want to understand not only the nature of the discs but the relationship between the composition of the polluted white dwarf atmospheres and the exoplanetary bodies they have accreted.

In order to interpret the absorption features observed in the atmospheres of white dwarfs, a good model for the structure of the white dwarf atmosphere and the diffusion of elements is required. Accreted elements are mixed within the white dwarf atmosphere and outer convective envelope, and diffuse out at the base of the convection zone. Recent improvements to ionisation models \citep{Heinonen2020} have highlighted their importance in calculating diffusion coefficients. Convective over-shooting has long been known to be important. Recent radiation-hydrodynamic simulations of white dwarf convection zones \citep{Tremblay2013,Tremblay2015} have shown that convective mixing zone could extend 2.5-3.5 pressure scale heights below the instability limit \citep{Cunningham2019}. The presence of metals in the envelope can themselves lead to sharp gradients in density, triggering instabilities such as the thermohaline (fingering) instability. This can lead to an under-prediction of accretion rates by orders of magnitude \citep{Deal_2013, Bauer2018, Bauer2019, Wachlin2017}, although its application to white dwarf convection zones has been questioned \citep{Koester_thermohaline}. It is, however, clear from the above papers that the thermohaline instability is not important for the cool, helium-rich white dwarfs considered in this work.

Interpretation of the accreted abundances is crucial. Thus far, most white dwarfs have accreted material dominated by Mg, Fe, Si, and O, possibly hinting that solar system-like geologies are common in the galaxy \citep{JuraYoung2014, Harrison2018}.
Evidence for the accretion of fragments of differentiated bodies onto white dwarfs has emerged due to the detection of pollutant bodies with high abundances, relative to the Solar System, in either the siderophilic elements or the lithophilic elements \citep{Zuckerman2011, Xu2013, Wilson2015}. In \cite{Harrison2018} it was suggested that for at least one system differentiation, collisions, and fragmentation of exoplanetary bodies is required to explain the atmospheric abundances. The accretion of differentiated fragments is expected, as not only do we find such bodies in the Solar System's meteorite suites \citep{WIIK1956, Scott1975}, but models of the collisional evolution of proto-planets predict a population of such fragments \citep{Marcus2009, Carter2015, Carter2017, bonsorleinhardt}.

An excess abundance of observed oxygen compared with that which could be sequestered in the form of metal oxides using the observed metal abundances is often used as evidence for the accretion of water ice \citep{Farihi2013,  Raddi2015, Xu2017}. \cite{Harrison2018} constrained the formation temperature of the accreting material, finding evidence for a wide spread in pollutant formation temperatures, ranging from dry volatile-poor asteroids to icy volatile-rich comet analogues. The accretion of water ice is not surprising as ice species at some orbital locations are expected to survive the post-main sequence evolution of the host star \citep{JuraXu2010, Malamud2016}.

The masses of the bodies which pollute white dwarfs are difficult to constrain. The mass of metals in the convective zone can be found using the spectral features and white dwarf atmospheric models \citep{Koester09, Koester2010}. \cite{Farihi2010} and \cite{Girven2012} found that the mass of metals in cool externally polluted DB white dwarf convective zones ranges from $10^{16}$ to $10^{22}$\,kg (using the  Ca abundance and assuming a bulk Earth composition for the pollutants). However, the mass of metals in the convective zone is not equivalent to the mass of the polluting body (bodies), as some of the mass may still be in the accretion disc and, depending on the length of time that the body has been accreting onto the white dwarf, some of the mass may have sunk out of the upper convective zone and no longer be observable. Under the assumption that the pollutant is a single body, the range of masses derived by \cite{Farihi10ism} and \cite{Girven2012} can be taken as lower limits on the mass of the planetary bodies accreted. Thus, it is currently hypothesised that the pollutants have masses similar to Solar System asteroids.

This work aims to test further the validity of the model outlined in \cite{Harrison2018} by using the large sample of white dwarf atmospheric abundances derived by \cite{Hollands2017} to constrain the origin of the pollutant bodies. We also aim to present a valuable upgrade to the model in \cite{Harrison2018} from frequentist to Bayesian, such that the model can now be applied to any newly discovered polluted white dwarfs, even where relatively few elements are detected. In Section \ref{methods} we outline the polluted white dwarf data analysed in this work, along with the methods used to constrain the formation history of the pollutants. In Section \ref{results} we present the results of this analysis and finally, in Section \ref{diss}, we discuss the caveats of our work and the implications of our results.

\section{Methods}\label{methods}

\subsection{The polluted white dwarf data set}\label{PWDD}

Our analysis focuses on the 230 externally polluted cool DZ white dwarfs presented in \cite{Hollands2017}. \cite{Hollands2017} derived atmospheric abundances of Ca, Mg, and Fe relative to He, and in some cases, Ti, Ni, Cr and Na abundances were also found. The aim of this work is to show the power of the probabilistic model presented here to provide conclusions when examining a large population of white dwarfs, even where relatively few elements are detected.

For this work we use the improved Mg abundances from \cite{Blouin2020}, who re-analysed the \cite{Hollands2017} systems using models that include the appropriate input physics for the fluid-like conditions of cool He-dominated white dwarf atmospheres. In particular, a detailed account of pressure ionization and non-ideal effects to the continuum opacities is included \citep{Blouin2018, Blouin2018B}, both of which affect the conditions in the line-forming region of the atmosphere. In addition, the Mg-b triplet ($\sim5170$\AA) is treated within the unified line broadening theory \citep{Allard2016}, instead of the more approximate \cite{Walkup1984} formalism. We note here that the magnetic systems are no longer considered in this analysis, nor is SDSSJ1055+3725, a suspected unresolved binary, leaving 202 systems. The abundances of Ca, Mg and Fe were re-fitted for all systems, including uncertainities. Except for the Mg abundances, the atmospheric parameters derived by \cite{Blouin2020} are largely consistent with those reported in \cite{Hollands2017}, which justifies this approach. The quoted errors include the uncertainty related to our imprecise knowledge of the stellar temperature, as well as the change in the abundance of one element, for a given stellar temperature. 
The elemental ratios for other species such as Cr/Ca, Ti/Ca, Ni/Ca or Na/Ca were assumed not to change as a result of the new models, and were used to calculate the new scaled abundances, listed in Table A1. This technique was also verified using test systems, as indicated by a $^\dagger$ in Table A1. Errors on Ti, Ni and Na were then calculated relative to He in dex by:
\begin{equation}\label{eq1}
\sigma_{X} = \sqrt{\sigma_{\textrm{systematic}}^{2} + \left(\frac{1.28}{\textrm{(S/N)}}\right)^{2}}
\end{equation}
where $\sigma_{\textrm{systematic}}$ is the assumed systematic error for the atmosphere elemental abundance relative to He in dex, (S/N) is the median signal to noise ratio of the spectra between 4500 and 5500\,\AA. For a handful of systems, re-analysis of the spectra placed in question the secure detection of Cr, Ni, Ti or Na. For these systems, no Cr, Ni, Ti or Na abundance was included in the analysis.

\subsection{A model to explain the composition of polluted white dwarf atmospheres} \label{WDCPWDA}

The model presented in this work is an extension of the model in \cite{Harrison2018}. We assume that each polluted white dwarf has accreted a single planetary body and that the observed compositions reflect the composition of this planetary body prior to accretion, modified only by the fact that different elements sink out of the observable atmosphere at different rates.  The potential differential sinking of different elements must be taken into account as the sample analysed in this work is composed of the old He dominated white dwarfs where sinking timescales can vary between tens of thousands of years and tens of millions of years. We follow \cite{Koester09} and the accretion is parameterised in terms of the time since the current accretion episode started and the total length of time for which the given accretion episode can continue, in order to take into account the possibility that accretion has now finished. 

The model presented finds the most likely pressure and temperature that could lead to the observed volatile elemental abundances. To do this it is assumed that the body is composed of material which condensed out of a nebula gas in chemical equilibrium and that the nebula gas could have a range of initial compositions. We consider that this range of initial conditions is likely to be similar to the range of initial material out of which stars formed and utilise a sample of nearby main-sequence stars as a proxy for these initial compositions. Assuming this simplistic model to be true and ignoring migration, or vertical mixing, these pressure-temperature conditions are equated to a location in a protoplanetary disc. In the model each accreted planetary body also has the potential to be a fragment of a larger body that differentiated to form a core, a mantle, and potentially a crust. The model determines the most likely core mass fraction (and potentially crustal fraction) of the fragment accreted by the white dwarf.

In the following sections, we outline further details of the model.

\subsubsection{The phase of accretion} \label{PAS}

Heavy elements present in the atmospheres of white dwarfs are expected to sink from the thin upper convective zone, due to the strong gravitational field of the white dwarf. Different chemical elements sink at different rates. Thus, the observed relative elemental abundances may not match those of the accreted material.

While accretion is occurring the elemental abundance ratio between two elements, A and B, in the atmosphere of the white dwarf are related to the elemental abundance ratio between the two elements in the pollutant by Equation \ref{eq2} \citep{Koester09}.
\begin{equation}\label{eq2}
\left(\frac{A}{B}\right)_{\textrm{O}} = \left(\frac{A}{B}\right)_{\textrm{P}}   \left (\frac{\tau_A}{\tau_B}\right) \frac{(1-e^{-t/\tau_A})}{  (1-e^{-t/\tau_B})} 
\end{equation}
\noindent where $\left(\frac{A}{B}\right)_{\textrm{O}}$ are the observed abundances of A and B, whilst $\left(\frac{A}{B}\right)_{\textrm{P}}$ are the abundances in the polluting material, $\tau_{\textrm{A}}$ and $\tau_{\textrm{B}}$ are the sinking timescales through the atmosphere of the white dwarf of elements A and B respectively and $t$ is the time passed since accretion started. From Equation \ref{eq2} we can see that initially the atmospheric abundances match those of the accreted material (the build-up phase) but as time passes, diffusion through the  atmosphere of the white dwarf and accretion onto the white dwarf equilibrate and the relationship between the atmospheric abundances and the abundances of the pollutant body tend towards a constant value (the steady state phase).

\begin{figure*}
	\includegraphics[width=2\columnwidth]{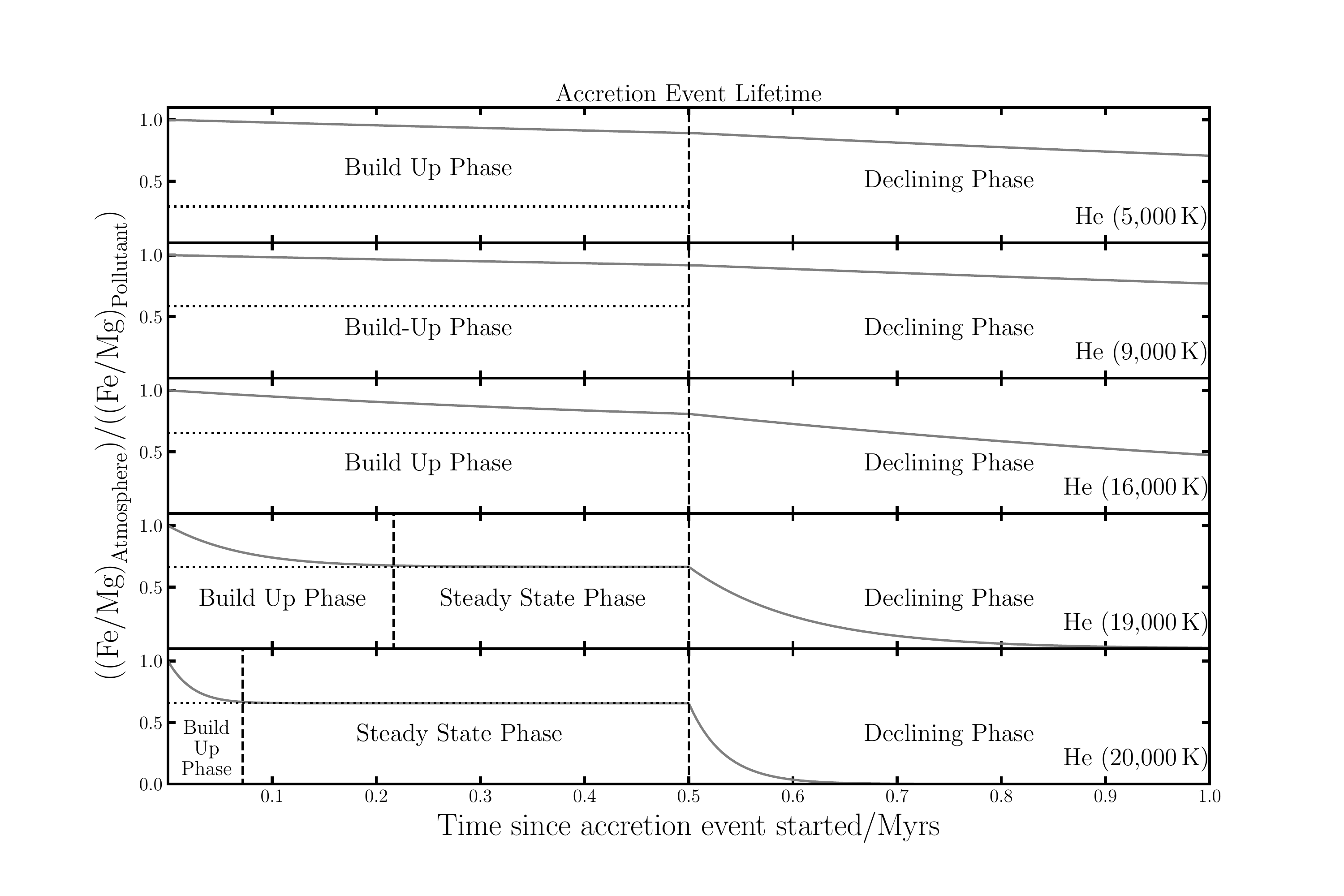}
 \caption{The abundance of Fe relative to Mg observed in a 0.6 solar mass white dwarf's helium dominated atmosphere as a function of time, for a fixed accretion event lifetime of 0.5\,Myrs, and for various white dwarf surface temperatures. Sinking occurs on shorter timescales in the hotter stars. }
    \label{fig:sink}
\end{figure*}

After accretion ceases the pollutant material will begin diffusing out of the upper convective zone of the white dwarf's atmosphere in a manner dependent on the individual elemental sinking timescales (the declining phase). The elemental abundance ratio between two elements, A and B, in the atmosphere of the white dwarf, $\left(\frac{A}{B}\right)_{\textrm{O}}$, are now related to the elemental abundance ratio between the two elements in the pollutant, $\left(\frac{A}{B}\right)_{\textrm{P}}$, by Equation \ref{eq3} \citep{Koester09}.

\begin{eqnarray}
\label{eq3}
\left(\frac{A}{B}\right)_{\textrm{O}} = \left(\frac{A}{B}\right)_{\textrm{P}}   \left (\frac{\tau_A}{\tau_B}\right) e^{\gamma(t-t_{\rm event})}  \frac{(1-e^{-t_{\rm event}/\tau_A})}{  (1-e^{-t_{\rm event}/\tau_B})} 
\end{eqnarray}
where:
\begin{equation}
    \gamma = \left ( \frac{\tau_B-\tau_A}{\tau_A\tau_B}\right)
    \end{equation}

\noindent where $t$ is the total time passed since accretion started, and $t_{\textrm{d}}$ is the lifetime of the accretion event. Figure \ref{fig:sink} displays how the abundance of Fe relative to Mg changes in the atmospheres of He dominated white dwarfs with various temperatures as time passes. The phase of accretion a system can reach is related to both the time since the accretion event started ($t$) and the accretion event lifetime ($t_{\textrm{d}}$). 

In this work we consider the possibility that each system could be in any phase of accretion (build-up, steady state, or declining). We do this using two parameters, the time since accretion started ($t$) and the total time for which accretion can occur, the accretion event lifetime ($t_{\textrm{event}}$). The conversion between the pollutant abundances and atmospheric abundances are described by Equation \ref{eq2} and Equation \ref{eq3} depending on which value is larger, the time passed since accretion started or the accretion event lifetime. The quality of the fit for each value of $t$ and $t_{\textrm{event}}$ can then be used to constrain not only the phase of accretion that a system may be in, but also the accretion event lifetime.

We calculate sinking timescales for all systems using linear interpolation on a grid of logarithms of simulated diffusion timescales, based on models presented in \citep{Koester2020}. For He-dominated white dwarfs, we interpolate on a newly calculated 3D grid. The presence of a one pressure scale height convective overshoot is assumed. The three variables we interpolate over are the white dwarf's temperature, $T_{\rm eff}$, surface gravity $log(g)$ and observed log(Ca/He). Rather than re-calculating the atmosphere model for every system with its exact abundances, instead bulk Earth abundances are assumed for the metals in the atmosphere, and the model with the nearest log (Ca/He) used. If a system lies outside the boundaries of the relevant grid, we linearly extrapolate to find its sinking timescales.

\subsubsection{The initial composition of the planetesimal forming disc} \label{ICPFD}

In our Solar System, chondritic meteorites have compositions which broadly match those of the Sun, taking into account depletion of volatiles. We assume that white dwarf pollutants, like the chondrites, formed in a protoplanetary disc around their host star. Not all protoplanetary discs will form from a nebula gas cloud of solar composition, and as the initial composition of the protoplanetary disc determines the possible compositions of planetesimals which form from the disc, variability in the nebula composition must be accounted for. In order to consider this effect we use the range of compositions seen in nearby stars as a proxy for the range of compositions likely to occur in exo-protoplanetary discs.

A sample of 1617 nearby FGK stars \citep{FischerBrewer2016} was chosen to be a suitable sample for the range of possible compositions expected. Nearby FGK stars were chosen in order to maintain the same approximate formation age as the polluted white dwarf progenitors (which were most likely earlier spectral types), thus, to first order the same diversity in chemical abundances. 

The sample of stars was refined to 958 stars as outlined in \cite{Harrison2018} and was then sorted by metallicity ([Fe/H]). We then took the host star metallicity index ($\textrm{[Fe/H]}_{\textrm{index}}$, the rank of the metallicity in the sample from 0 (the lowest metallicity) to 957 (the highest metallicity)) as a free parameter allowing us to model the variation in initial nebula compositions in the stellar sample using only one parameter.

\subsubsection{The heating which takes place during formation} \label{HDFSE}

Most pollutants of white dwarfs are depleted in the volatile elements, and many white dwarf pollutants show trends related to elemental volatility \citep{JuraYoung2014, Dufour2016,Harrison2018}. Such trends are indicative of the temperature experienced by the material either during formation or afterwards. Therefore, in order to model the composition of planetesimals, the temperature they experience should be taken into account.

In this work we use the simplest possible approximation and assume that the observed trends related to volatility developed during formation in chemical equilibrium, and therefore we can employ a Gibbs free energy minimisation model which links a the pressure-temperature space to a solid state composition. To model the expected primitive rocky planetesimal abundances we use the stellar elemental abundances from \cite{FischerBrewer2016}, as mentioned in Section \ref{ICPFD}, as inputs for the program HSC chemistry version 8, a Gibbs free energy minimisation solver, and calculate the abundances of the solid species that condensed out of the gaseous nebula over the pressure-temperature space mapped out by the analytic disc model derived in \cite{Chambers2009}. The chosen disc model is an irradiated viscous disc model with an alpha parameterisation which models the evolving pressure-temperature space in an irradiated viscously heated protoplanetary disc around an A0 type star. The full specification of the viscous irradiated protoplanetary disc model and the equilibrium chemistry model are presented in \cite{Harrison2018}. For a discussion of the validity of our models and assumptions see Section \ref{disscav}. It is important to note that in this work we have modified the input parameters in the disc model to match those of an A0 type star rather than a solar type star as in \cite{Harrison2018}. This is because it is expected that the progenitors of the white dwarfs in the \cite{Hollands2017} sample were most likely A0 type main sequence stars \citep{Veras2016}. The parameters were modified from the values used in \cite{Chambers2009} using the stellar evolution curves in \cite{Siess2000} and the stellar mass protoplanetary disc mass relation given in \cite{Andrews2013}. 

The compositions of primitive planetesimals are, therefore, modelled to be functions of the metallicity of the host star ($\textrm{[Fe/H]}_{\textrm{index}}$), the formation distance from the host star ($d_{\textrm{\,formation}}$), planetesimal formation time ($t_{\textrm{\,formation}}$) and the feeding zone size from which the planetesimal accretes material ($z_{\textrm{\,formation}}$). The quality of the fit for a given $d_{\textrm{\,formation}}$, and $z_{\textrm{\,formation}}$ can then be used to constrain the formation temperature of the accreted material.

\subsubsection{Differentiation, collisions, and fragmentation} \label{DCF}

Many pollutants of white dwarfs show enhancement or depletion in siderophile (iron-loving) and/or lithophile (silicate-loving) elements in comparison to bulk Earth \citep{JuraYoung2014}. Collisions during planet formation and in the protoplanetary disc phase are common and often disruptive \citep{Vries2016}. Disruptive collisions between planetary bodies that have differentiated can lead to the creation of fragments with non-primitive abundances \citep{Marcus2009, bonsorleinhardt, Carter2015}. Therefore, differentiation, collisions, and fragmentation should be modelled in order to reproduce the full suite of expected planetesimal compositions.

To model the expected non-primitive planetesimal chemical abundances present in extrasolar systems we use the simple Earth-based differentiation model outlined in \cite{Harrison2018}. The model calculates the expected compositions of the cores, mantles, and crusts of the planetesimals modelled in Section \ref{HDFSE} by allowing certain elements to preferentially move into the core and crust of the planetary body during the differentiation process. The compositions of the non-primitive planetesimals produced by collisions can then be obtained by linearly combining various amounts of core-like, mantle-like and crust-like material. Equation \ref{eq4} outlines how enhanced an individual element, X, becomes in the collisional fragment relative to a original parent planetesimal. 
\begin{equation}\label{eq4}
\resizebox{.92\hsize}{!}{$ E_{\textrm{x}}X_{\oplus} = f_{\textrm{o}}X_{\textrm{o}\oplus}+ f_{\textrm{c}}X_{\textrm{c}\oplus} + \left(\frac{f_{\textrm{m}}}{N_{\textrm{m}}}\right)(X_{\oplus}-N_{\textrm{o}}X_{\textrm{o}\oplus}-N_{\textrm{c}}X_{\textrm{c}\oplus}) $}
\end{equation}
\noindent where $ E_{\textrm{x}}$ is the enhancement factor of element X, $X_{\oplus}$ is the abundance of element X in bulk Earth, $X_{\textrm{o}\oplus}$ is the abundance of element X in the Earth's oceanic crust, $X_{\textrm{c}\oplus}$  is the abundance of element X in the Earth's core, $f_{\textrm{o}}$ is the fraction of the fragment which is made up of oceanic crust-like material, $f_{\textrm{m}}$ is the fraction of the fragment which is made up of mantle-like material, $f_{\textrm{c}}$ is the fraction of the fragment which is made up of core-like material, $N_{\textrm{o}}$ is the fraction of the parent body which is made up of oceanic crust-like material, $N_{\textrm{m}}$ is the fraction of the parent body which is made up of mantle-like material, and $N_{\textrm{c}}$ is the fraction of the parent body which is made up of core-like material. All elemental abundances for the Earth are taken from \citep{McDonough2003, White2014}. Throughout the text we discuss number fractions and abundances by number, rather than mass fractions. We fix the parent core and crust fractions to be those of planet Earth ($N_c=0.17$ and $N_o=0.01$). In \cite{Harrison2018}, these were considered to be free parameters, but for the present sample there is insufficient information to warrant these additional parameters. The full specification of the differentiation model is presented in \cite{Harrison2018}.


The compositions of non-primitive planetesimals are, therefore, modelled to be functions of 
the fragment core fraction ($f_{\textrm{c}}$) and the fragment crust fraction ($f_{\textrm{o}}$), as well as metallicity of the host star ($\textrm{[Fe/H]}_{\textrm{index}}$), the formation time ($t_{\textrm{\,formation}}$), the formation distance from the host star ($d_{\textrm{\,formation}}$), and the feeding zone size ($z_{\textrm{\,formation}}$). The quality of the fit for a given $f_{\textrm{c}}$, and $f_{\textrm{o}}$ can then be used to constrain the geological and collisional history of the accreted material. 

\subsubsection{Converting planetesimal abundances to atmospheric abundances}

In order to compare the atmospheric abundances of the polluted white dwarfs to our modelled planetesimal abundances, we must know how much of the convective zone is composed of pollutant material. We avoid any assumptions by allowing the pollutant to convective zone number fraction to be a free parameter in the model ($P_{\textrm{\,fraction}}$). This is the sum of the total observed pollutants, rather than all polluting material. This is converted to a total pollutant mass for the white dwarf system using the convective zone mass for a star of that surface gravity, temperature and Ca abundance (see \S\ref{PAS}). This information can be combined with the knowledge derived from the phase of accretion in order to produce a constraint on the total pollutant mass that has ever been accreted onto the star during the accretion event which is currently being observed. 

\begin{figure*}
	\includegraphics[width=\textwidth]{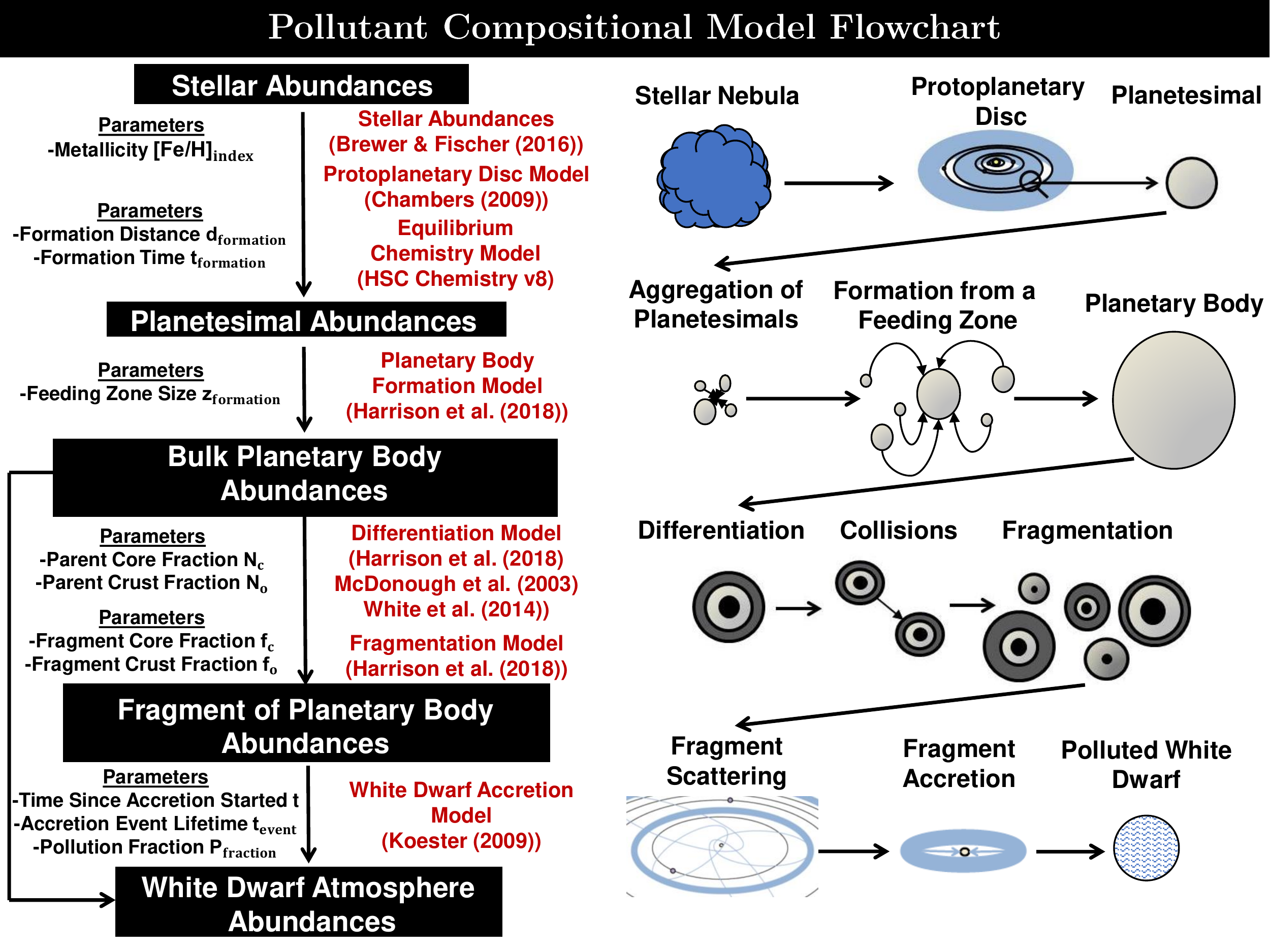}
 \caption{A flowchart summarising the model used and the parameters involved when modelling the composition of white dwarf pollutants.}
    \label{fig:1}
\end{figure*}

Figure \ref{fig:1} displays a flowchart which summarises the model used in this work, highlighting the 11 free parameters in the full model: stellar metallicity ($\textrm{[Fe/H]}_{\textrm{index}}$), planetesimal formation distance ($d_{\textrm{\,formation}}$), planetesimal feeding zone size ($z_{\textrm{\,formation}}$),  fragment core fraction ($f_{\textrm{c}}$), fragment crust fraction ($f_{\textrm{o}}$), time since accretion event started ($t$), accretion event lifetime ($t_{\textrm{event}}$), and white dwarf atmospheric pollution fraction ($P_{\textrm{fraction}}$).

\subsection{Statistically constraining the origin of pollutants} \label{SCOP}

\subsubsection{Bayesian inference via nested sampling} 

In order to select the most probable origin of the pollutant material and rule out areas of parameter space within our models which cannot accurately reproduce the observed atmospheric abundances in the individual white dwarfs, we use the nested sampling algorithm MultiNest \citep{Feroz2008, Feroz2009, Feroz_2019} via the python package PyMultiNest \citep{MultiNest}. 

The validity of a given model relative to another is assessed using the Bayesian evidence ($\mathcal{Z}$). This is defined as the product of the assumed prior distribution for each parameter ($\pi$) and the likelihood function ($\mathcal{L}$). The posterior probability distribution for each parameter ($p(\theta|\textrm{\textbf{X}}_{\textrm{obs}}, M_{i})$) is used to constrain the potential values of each parameter, given the observations. The ratio of the Bayesian evidence (Bayes factor) is used to compare models.

MultiNest calculates the Bayesian evidence of a model ($\mathcal{Z}$) by numerically integrating the product of the assumed prior distribution for each parameter ($\pi$) and the likelihood function ($\mathcal{L}$)

\begin{equation} \label{Evidence}
\resizebox{.7\hsize}{!}{$\mathcal{Z}(\textrm{\textbf{X}}_{\textrm{obs}}|M_{i}) = \int_{\theta}\,\, \mathcal{L}(\textrm{\textbf{X}}_{\textrm{obs}}|$\boldmath$\theta$\unboldmath$,M_{i})\, \pi($\boldmath$\theta$\unboldmath$|M_{i})\,\, d\theta  $}
\end{equation}

\noindent where $\textrm{\textbf{X}}_{\textrm{obs}}$ is a vector containing the abundances observed in the polluted white dwarf's atmosphere, $M_{i}$ is the model chosen to explain the atmospheric abundances, and \boldmath$\theta$\unboldmath\, is a vector containing the set of parameters included in the chosen model. The likelihood ($\mathcal{L}$) is defined by 

\begin{equation} \label{Likelihood}
\resizebox{.9\hsize}{!}{$\textrm{ln}(\mathcal{L}(\textrm{\textbf{X}}_{\textrm{obs}}|\theta,M_{i})) = -\frac{1}{2} \sum\limits_{X} ^{\rm{Ca,Fe,...}}\left({\frac{\left( X_{\rm{obs}} \,-\,  X_{\rm{mod}}(\theta)\right)} {\sigma_{X}^{2}}^{2}} + \textrm{ln}(2 \pi \sigma_{X}^{2})\right) $}
\end{equation}

\noindent where the sum is over all the observed elemental abundances, $X_{\rm{mod}}$(\boldmath$\theta$\unboldmath) is the modelled abundance for a given set of parameters \boldmath$\theta$\unboldmath, and $\sigma_{X}$ is the error on the individual observed elemental abundances. The prior ($ \pi($\boldmath$\theta$\unboldmath$|M_{i})$) for each parameter is taken to be a uniform (or log-uniform) distribution between an upper and a lower bound, the priors assumed for each parameter are listed in Table \ref{Priors} and discussed in Section \ref{CoP}.

The Bayesian evidence is efficiently calculated by MultiNest by exploring progressively increasing iso-likelihood contours in the parameter space such that the iso-likelihood contours contain a set of live points taken from incrementally shrinking ellipsoids. We run the algorithm with 2,000 live points as we found that this was sufficient to produce errors on the log evidence of the order 0.10, while also minimising the run time of the code.

The posterior probability distribution ($ p($\boldmath$\theta$\unboldmath$|\textrm{\textbf{X}}_{\textrm{obs}}, M_{i})$) for each parameter is found as a convenient byproduct of the evidence calculation as the posterior is related to the likelihood, evidence, and prior by:

\begin{equation} \label{Posterior}
\resizebox{.6\hsize}{!}{$ p(\theta|\textrm{\textbf{X}}_{\textrm{obs}}, M_{i}) = \frac{\mathcal{L}(\textrm{\textbf{X}}_{\textrm{obs}}|\theta,M_{i})\, \pi(\theta|M_{i})}{\mathcal{Z}(\textrm{\textbf{X}}_{\textrm{obs}}|M_{i})} $}
\end{equation}

\noindent The updated posterior distributions provide constraints on each parameter given the observations, and thus, can be used to constrain the origin of the pollutant material. In this work for each system we can potentially find constraints on each of the 8 parameters in our full model ($\textrm{[Fe/H]}_{\textrm{index}}$, $t$, $t_{\textrm{event}}$, $d_{\textrm{\,formation}}$, $z_{\textrm{\,formation}}$, $f_{\textrm{c}}$, $f_{\textrm{o}}$, and $P_{\textrm{\,fraction}}$).

In order to compare models, find the optimum model to explain a systems abundances, and estimate the statistical significance of various pollutant histories (for example, to what significance the white dwarf requires the accretion of a fragment of a differentiated body) a Bayes factor ($\mathcal{B}_{ij}$) can be used. The Bayes factor between two models ($i$ and $j$) is calculated using:

\begin{equation} \label{BayesFactor}
\resizebox{.3\hsize}{!}{$\mathcal{B}_{ij} = \frac{\mathcal{Z}(\textrm{\textbf{X}}_{\textrm{obs}}|M_{i})}{\mathcal{Z}(\textrm{\textbf{X}}_{\textrm{obs}}|M_{j})}  $}
\end{equation}

\noindent where, as before, $\mathcal{Z}$ is the Bayesian evidence, $\textrm{\textbf{X}}_{\textrm{obs}}$ is a vector containing the abundances observed in the polluted white dwarf's atmosphere and $M_{i}$ and $M_{j}$ are the models chosen to explain the atmospheric abundances. 

The Bayes factor between two models can be converted into a sigma-significance of a given additional parameter using Equation \ref{sigB} \citep{Sellke2001}. 

\begin{equation} \label{sigB}
\sigma =  \sqrt{2}\,\textrm{erfcinv}\left( \Re\left(e^{W\left(-\frac{1}{e\mathcal{B}_{ij}}\right)}\right) \right)
\end{equation}
\noindent where erfcinv is the inverse of the function shown in Equation \ref{erfc}, 

\begin{equation} \label{erfc}
\textrm{erfc}(x) =  \frac{2}{\sqrt{\pi}} \int_{x}^{\infty} e^{-t^{2}} dt 
\end{equation}

\noindent and W is the Lambert W function whose defining equation is Equation \ref{LW},

\begin{equation} \label{LW}
z =  W(z)e^{W(z)}
\end{equation}
\noindent where z is a complex number.

In this work we concentrate on the Bayes factors, and thus the sigma-significance, which indicate the requirement for the inclusion of non-build-up phase accretion, core-mantle differentiation, core-mantle-crust differentiation, and heating during formation. We also derive and report the $\chi^{2}$ per element observed values for each best-fit model. This is because it is important to check that the best-fit models (the ones with the highest Bayesian evidence) are good fits to the data before drawing any sigma-significance conclusions.

For a more detailed discussion of the MultiNest algorithm which we use in this work to provide fast and robust parameter estimation and model comparisons see \cite{BennekeSeager2013}.

\subsubsection{The choice of prior parameter distributions}\label{CoP}

\begin{table}
	\centering
	\caption{Prior distributions assumed for the 8 parameters in our full model, where $Z_{\rm tot}$ is the sum of all observed pollutants.}
	\label{Priors}
\begin{tabular}{ c c c c }
\hline
Parameter & Prior & Range  \\
\hline
$\textrm{[Fe/H]}_{\textrm{index}}$ & Uniform & 0 to 957 \\
$d_{\textrm{\,formation}}$ & Log-uniform & $10^{-2}$ to $R_{\textrm{disc}}$($t_{\textrm{\,formation}}$)\,AU \\
$z_{\textrm{\,formation}}$ & Uniform & 0 to 0.15\,AU \\
$f_{\textrm{c}}$ & Uniform & 0 to 1 \\
$f_{\textrm{o}}$ & Uniform & 0 to 1-$f_{c}$ \\
$t$ & Uniform & 0 to $t_{\textrm{event}}$ + 12\,$\tau_{\textrm{Mg}}$ \\
$t_{\textrm{event}}$ & Log-uniform & $10^{-6}$ to $10^{2}$\,Myrs \\
$P_{\textrm{fraction}}$ & Log-uniform & $10^{-1.5}S$ to $10^{+1.5}S$, $S =\textrm{log}(\textrm{Z}_{\rm tot}/\textrm{He})$
\\ \hline
\end{tabular}
\end{table}

As highlighted in the previous section, in order to perform Bayesian analysis, and calculate the posterior distribution for each free parameter defined in the model, prior distributions defined over a suitable range must be assumed for each parameter. The aim is to choose priors that cover the full potential parameter space available to the models, such that they do not influence the model comparison. Table \ref{Priors} contains the prior distributions assumed in this work.

The prior on $\textrm{[Fe/H]}_{\textrm{index}}$ is trivial as it is simply a uniform prior over the refined stellar compositional catalogue outlined in Section \ref{ICPFD}. Therefore, this prior only forces the initial composition of the protoplanetary disc to be one which is in the catalogue.

In order to model the fact that planetesimals can be composed of material which formed at a range of radii, a feeding zone parameter is included in the model ($z_{\textrm{formation}}$). Accretion is a stochastic process, migration and scattering are likely to be important and, therefore, a physically reasonable value of this parameter is difficult to obtain. Instead, we chose a uniform prior distribution for the feeding zone size between 0 and 0.15\,AU as we found that this was not unphysically large nor did it artificially constrict the model.

The priors on the fragment core and crust number fractions are outlined in Table \ref{Priors}. Additionally, we also do not allow the collisional fragments to be simultaneously enhanced in core-like and crust-like material, nor do we allow mantle-like material to be removed from a body without first removing the crust-like material (This alters the prior range and distribution for $f_{c}$ and $f_{o}$ given in Table \ref{Priors} to a non-trivial form). This allows improbable fragment compositions to be ignored and corresponds to the expected range of fragments formed from protoplanetary collisions \citep{Carter2017}. 

The prior on the pollution fraction was chosen to be log-uniform over a range $10^{-1.5}S$ to $10^{+1.5}S$, where $S = \log(Z_{\rm tot}/\textrm{He})$. Such a range was found to be sufficiently broad to model all the systems analysed in this work.

It is unclear how long each individual white dwarf accretion event lasts. Estimates of such event lifetimes range from tens of years \citep{Wyatt2014} to millions of years \citep{Girven2012}. Theoretical considerations of Poynting-Robertson drag driven accretion suggest that disc lifetimes for discs with a mass of $\sim10^{19}$\,kg are of the order of millions of years \citep{rafikov2, Rafikov1}. In this work we chose a log-uniform prior distribution of accretion event lifetimes ($t_{\textrm{event}}$) ranging from 1 year to 100 million years in order to capture the full extent of possible disc lifetimes.

 The prior distribution of the times since accretion started ($t$) are uniformly distributed between 0\,Myrs and 12\,$\tau_{\textrm{Mg}}$ after the accretion event has finished, where $\tau_{\textrm{Mg}}$ is the sinking timescale for Mg. The upper cut off was found to probe sufficiently far into the declining phase to model all the systems in the \cite{Hollands2017} sample.
 
As outlined above it is expected that all the priors chosen are non-informative, however, a full discussion of their validity will be presented in Section \ref{disscav}.

\section{Results} \label{results}

The 202 polluted white dwarf atmospheric abundances outlined in Section \ref{PWDD} were fitted using the model described in Section \ref{WDCPWDA} using the Bayesian analysis techniques presented in Section \ref{SCOP}. For each system the Bayesian evidence and $\chi^{2}$ per element was found for various combinations of model parameters as described by Table~\ref{ModelSummary}. Only models with the combinations of free parameters outlined in Table~\ref{ModelSummary} were used in order to minimise run time, while simultaneously robustly investigating whether each main model parameter was necessary. Table \ref{ModelSummary} summarises the free parameters of each model used in this work and the best model is defined as the model which has the highest Bayesian evidence. Once the various models have been used to fit the data, the model with the highest Bayesian evidence is selected as the optimum model. The $\chi^{2}$ per element of the model fit is then checked in order to confirm that the 'best' model is not simply the best out of a group of models which all struggle to reproduce the observations. In this work a model with a $\chi^{2}$ per element of less than 1 was considered to be a sufficient fit. By comparing the Bayesian evidence after removing or adding extra parameters a constraint on the necessity for the inclusion of such a parameter can be found. In this work we concentrate on the requirement for the inclusion of non-build-up phase accretion, core-mantle differentiation, core-mantle-crust differentiation, and heating during formation. Table A2 indicates the model with the highest Bayesian evidence for each system, whilst Table A3 includes the statistical significance for each system, alongside the best-fit model parameters. 


\begin{table*}
	\centering
	\caption{A table describing the models used in this work, in particular the free parameters considered for each model. Although models with a range of formation distances (feeding zone) and core-mantle or crustal differentiation were considered, they were not required for any system in this work. }
	\label{ModelSummary}
\begin{tabular}{c c c}
\hline
Model Label & Model Parameters & Model Description \\ 
\hline
M1 & $\textrm{[Fe/H]}_{\textrm{index}}$, $P_{\textrm{fraction}}$, $t_{\textrm{event}}$, $t$ &  Primitive planetesimal with no heating\\
M2 & $\textrm{[Fe/H]}_{\textrm{index}}$, $P_{\textrm{fraction}}$, $t_{\textrm{event}}$, $t$, $d_{\textrm{\,formation}}$ & Primitive planetesimal\\
M3 & $\textrm{[Fe/H]}_{\textrm{index}}$, $P_{\textrm{fraction}}$, $t_{\textrm{event}}$, $t$, $d_{\textrm{\,formation}}$, $z_{\textrm{\,formation}}$ & Primitive planetesimal with feeding zone\\
M4 &  $\textrm{[Fe/H]}_{\textrm{index}}$, $P_{\textrm{fraction}}$, $t_{\textrm{event}}$, $t$,  $f_{\textrm{c}}$  & Core-mantle differentiated fragment \\
M5 & $\textrm{[Fe/H]}_{\textrm{index}}$, $P_{\textrm{fraction}}$, $t_{\textrm{event}}$, $t$,  $f_{\textrm{c}}$, $f_{\textrm{o}}$ & Core-mantle-crust differentiated fragment \\
M6 &  $\textrm{[Fe/H]}_{\textrm{index}}$, $P_{\textrm{fraction}}$, $t_{\textrm{event}}$, $t$, $d_{\textrm{\,formation}}$, $f_{\textrm{c}}$  & Core-mantle differentiated fragment with heating \\

\hline
\end{tabular}
\end{table*}

\subsection{Evidence for heating during planet formation}

\begin{figure}
    \centering
    \includegraphics[width=1.15\columnwidth]{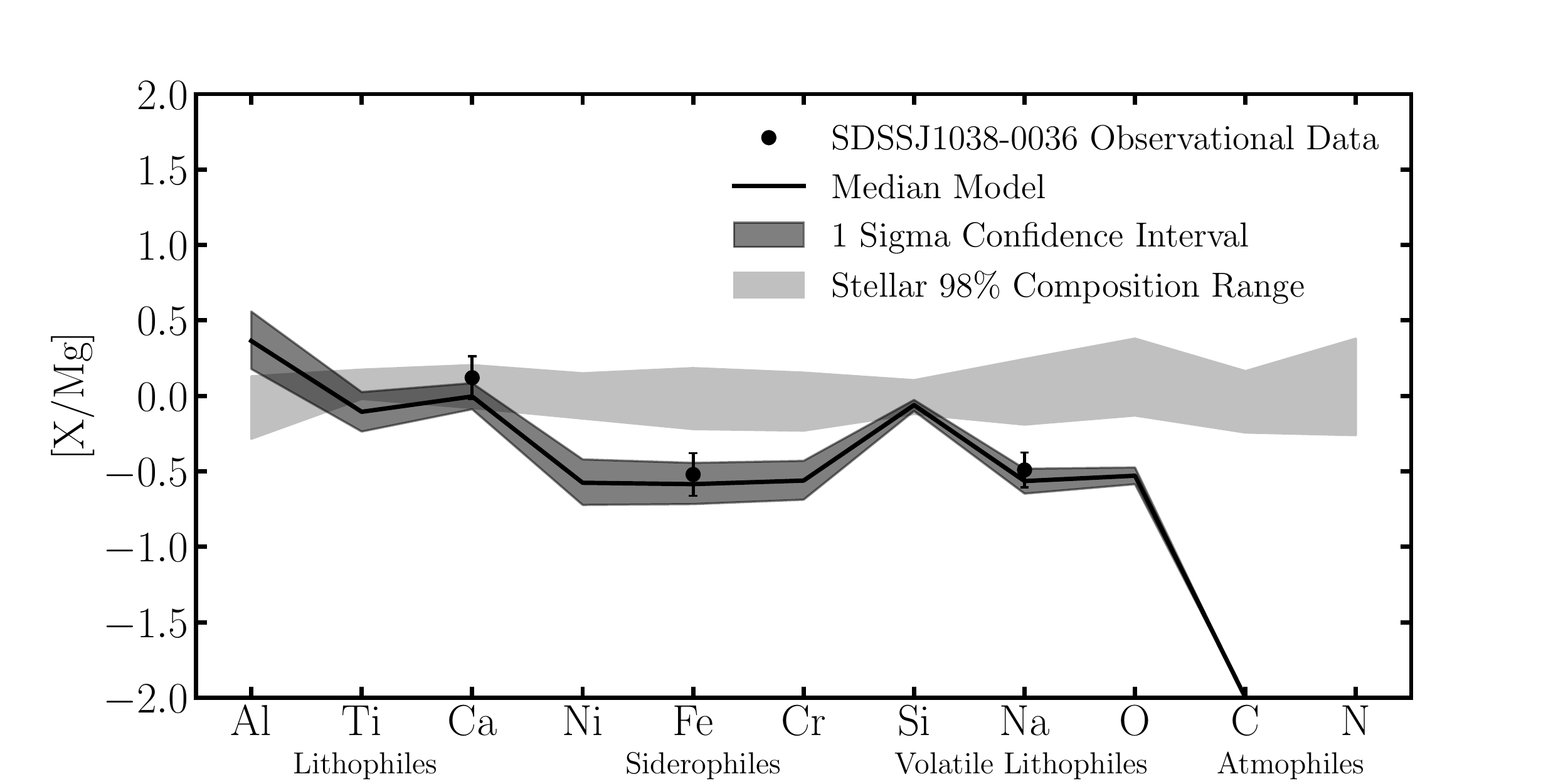}
    \caption{Volatile Depleted: Elemental ratios relative to Mg, normalised to solar, for the planetary material accreted by SDSS J1038-0036, where Na is depleted relative to Mg and Fe. The best-fit model suggests that this system is accreting in the declining phase, to explain the low Fe/Mg ratio. } 
    \label{fig:nadepleted}
\end{figure}

\begin{figure}
    \centering
    \includegraphics[width=\columnwidth]{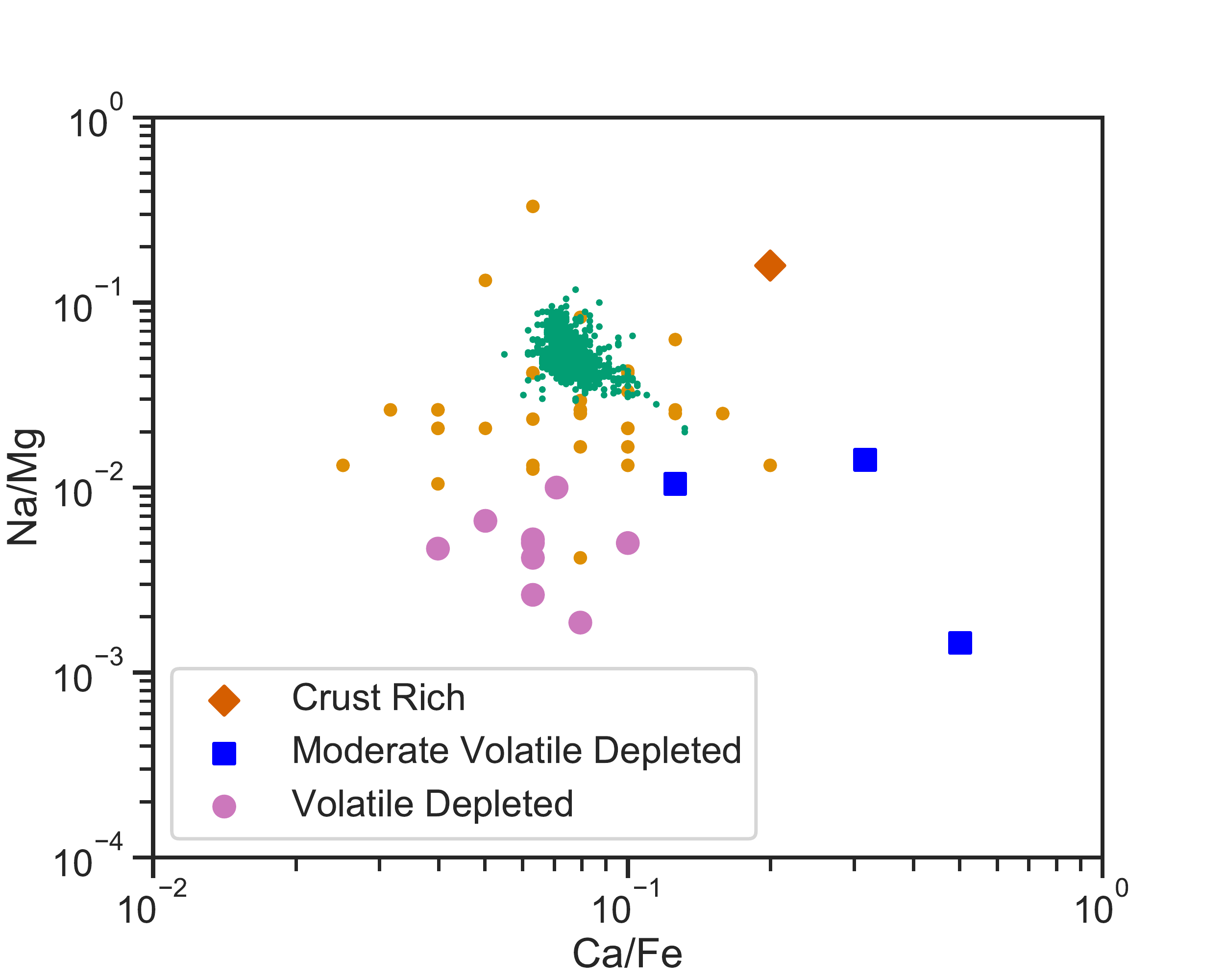}
    \caption{The Na/Mg and Ca/Fe ratio in the abundances of the material accreted by all systems. Those systems with low Na/Mg, where the best model suggests evidence of depletion of Na are shown as purple circles, and depletion of moderate volatiles (blue squares) whilst those with high Na/Mg found to be crust-rich are shown by the orange diamond. The stars in the \citet{FischerBrewer2016} sample are shown in light green. }
    \label{fig:na}
\end{figure}

\begin{figure}
	\includegraphics[width=\columnwidth]{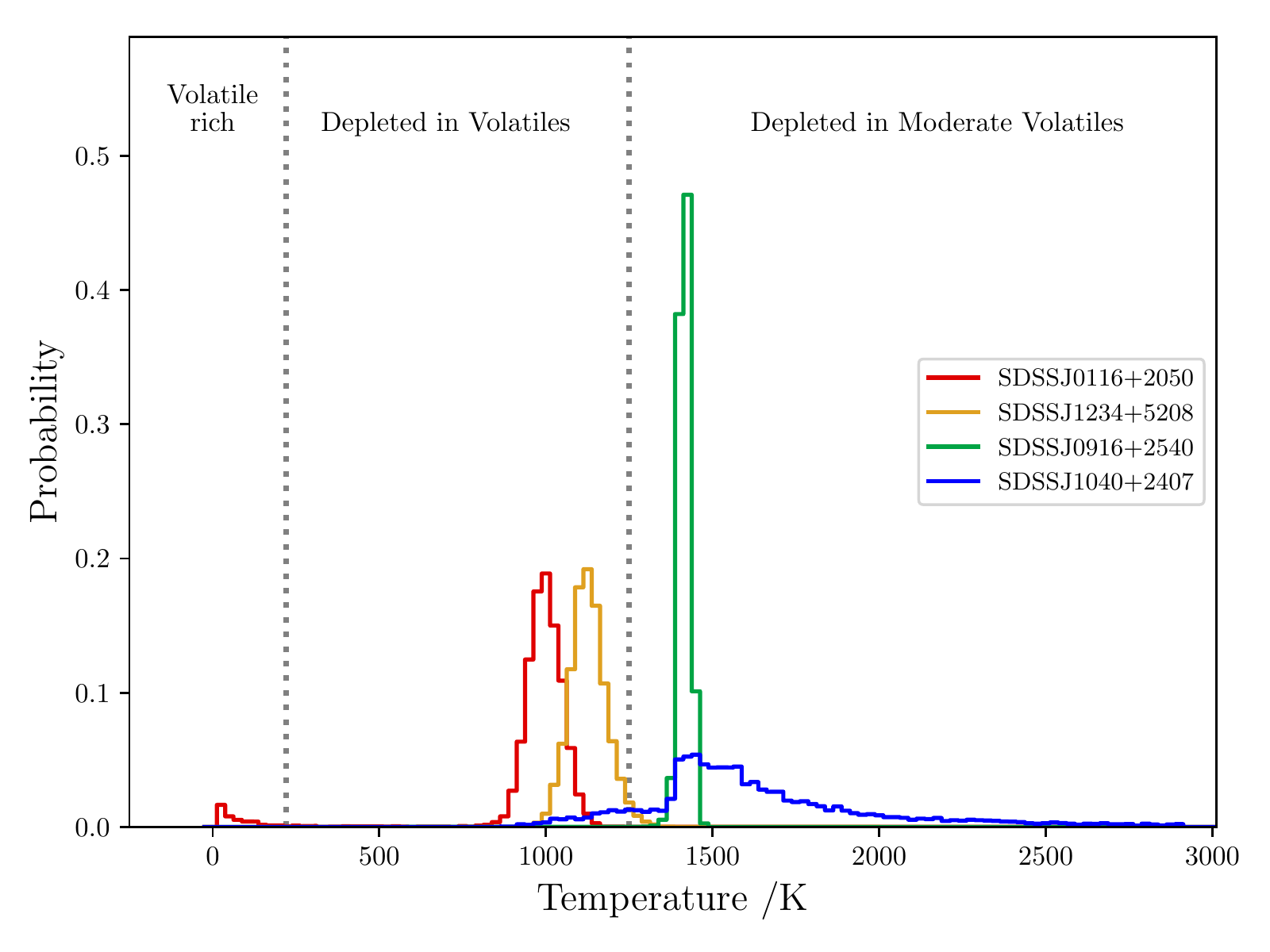}
 \caption{The posterior distributions of formation temperature for four example systems. Both SDSSJ0916+2540 and SDSSJ1040+2407 show evidence of being depleted in moderate volatiles (or rich in refractories such as Ca). For SDSSJ0916+2540 (green) the best-fit includes material from one location only, whilst for SDSSJ 1040+2407 (blue) the best-fit model includes a feeding zone, where material from a range of temperatures are required to explain the observed abundances. Both SDSSJ0116+2050 (red) and SDSSJ1234+5208 (yellow) are depleted in Na only, and their posterior distributions indicate formation temperatures below 1,400K. }
    \label{fig:temppost}
\end{figure}

All rocky planetary bodies are depleted in volatiles. Whilst, for the polluted white dwarfs considered in this work, we are unable to investigate the lack of very volatile species such as C, N or O and we cannot find evidence for icy planetesimals, we are able to conclude that some systems have accreted planetary bodies that are depleted in volatiles. Here we focus on two key signatures of volatile depletion. The first is the depletion of Na relative to other refractory species e.g. Mg, Ca. This can only be probed in the  systems where Na is detected, of which 12/50 show signs of volatile depletion, whilst 38 did not experience sufficient heating to deplete Na. Considering the incomplete condensation of the nebula gas, this means that 38/50 systems show no evidence of heating to temperatures above 1,000\,K. Fig.~\ref{fig:na} shows the Na/Mg ratios, compared to the Ca/Fe ratios for those 50
systems, alongside the model with the highest Bayesian evidence. An example system is shown in Fig.~\ref{fig:nadepleted}. The depletion of Na/Mg is very difficult to confuse with sinking effects, as Na sinks at a very similar rate to Mg. Both Na and Mg are lithophile and thus, Na/Mg is not significantly altered by core-mantle differentiation. We name these systems Volatile Depleted.  Fig.~\ref{fig:temppost} shows the posterior distributions for the temperature for systems that are depleted in volatiles are peaked at around 1,000K.

\begin{figure}
    \centering
    \includegraphics[width=1.15\columnwidth]{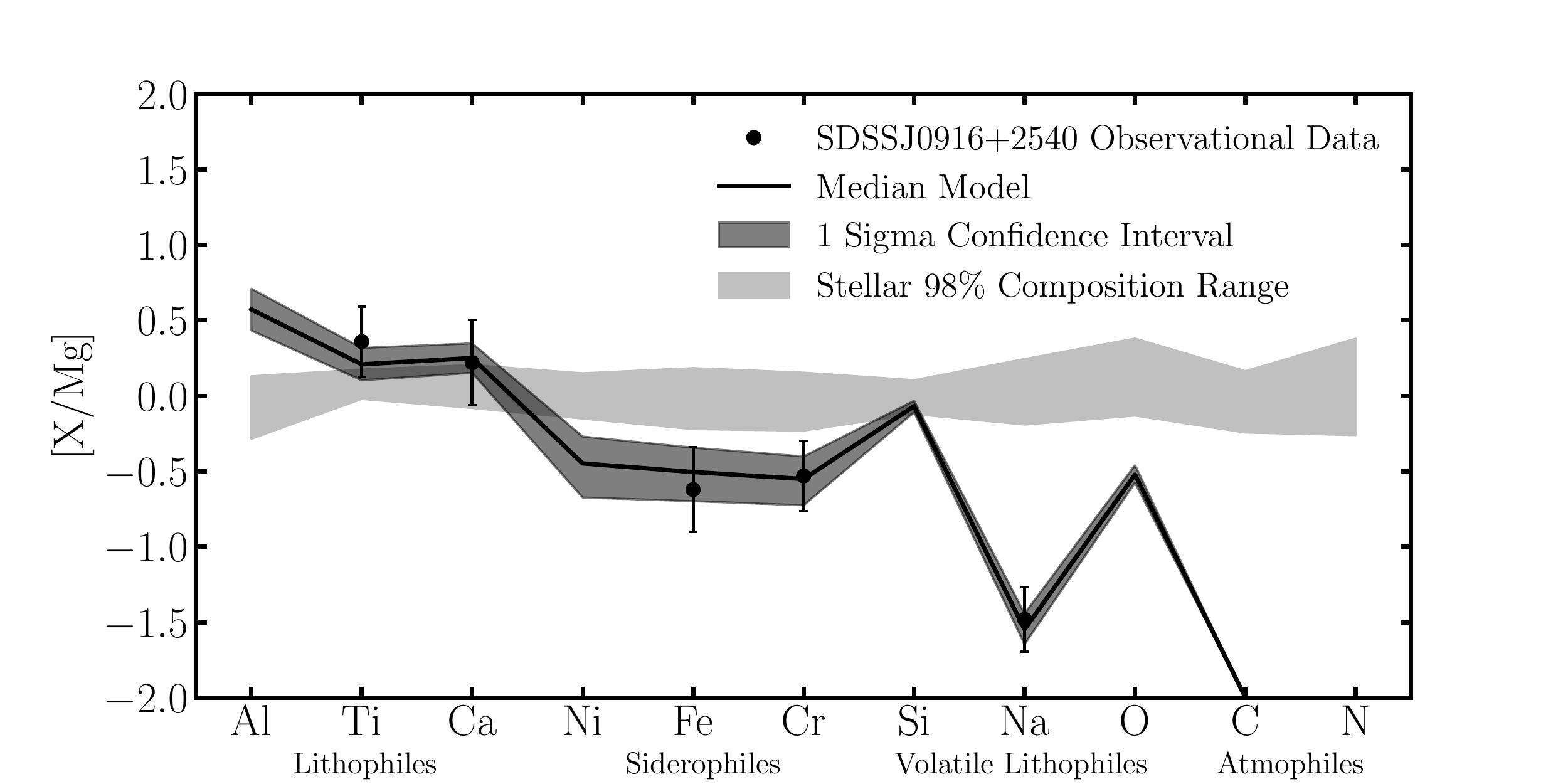}
    \caption{Moderate Volatile Depletion: In the same manner as Fig.~\ref{fig:nadepleted}, elemental ratios relative to Mg for the planetary material accreted by SDSS J0916+2540, where Ca and Ti are enhanced relative to Mg and Na is depleted. If this signature was caused by the incomplete condensation of the nebula gas, the planetary material experienced temperatures above 1,400K during formation. This system is found to be accreting in the build-up phase. }
    \label{fig:moderatevolatiledepleted}
\end{figure}

For those systems that have experienced even higher temperatures, moderately volatile species are depleted. Ca, Al and Ti are highly refractory species. For temperatures above 1,400\,K, iron and magnesium-rich minerals can be depleted compared to highly refractory calcium or aluminium rich minerals.  This can be seen as high Ca/Fe or Ca/Mg ratios in the accreted planetary material. Both high Ca/Fe and high Ca/Mg provide a signature that is hard to confuse with either core-mantle or crustal differentation or sinking. When Na is detected, this provides further evidence in support of the high temperatures experienced by the accreted planetary bodies. Fig.~\ref{fig:moderatevolatiledepleted} shows an example system, SDSS J0916+2540, where both Ca and Ti are enhanced relative to Fe, or Mg, and Na is depleted. Fig.~\ref{fig:temppost} shows the posterior distributions for the temperature for this system (green), which requires temperatures above 1,400\,K to explain its abundances. Also shown on Fig.~\ref{fig:temppost} is the posterior distribution for SDSSJ1040+2407, which can be best explained by material that has experienced a range of formation temperatures. 

\subsection{Phases of Accretion} 
Both the time at which the system is observed and the total length of time for which accretion (is) was on-going ($t_{\rm event}$) are unknown. Some systems are observed soon after accretion starts (compared to the sinking timescales of Myrs) such that atmospheric compositions are unaltered from the accreted material (build-up phase). Some systems are observed in a steady-state between accretion and diffusion (for times longer than the sinking timescale and shorter than the accretion event lifetime). Some systems are observed after accretion has finished (declining phase). The model finds the most likely phase of accretion, however, in many cases this is degenerate, with insufficient information available in the observed abundances to distinguish between accretion phases.

For those systems with only Ca, Mg and Fe detections, both Mg/Fe and Ca/Fe appear higher in the observed abundances due to the effects of sinking. For Mg/Fe, after one sinking timescale for Fe, this is by a factor of $\sim 2$, whilst for Ca/Fe this is by a factor of $\sim 1.2$. We note here, however, that these ratios come directly from our choice of model for the white dwarf atmosphere and diffusion (see Sec~\ref{PAS}). Fig.~\ref{fig:mg_fe_ca_fe} indicates the observed Mg/Fe ratios against Ca/Fe ratios, labelled with the most likely phase of accretion, as determined by the Bayesian model. In order to explain those systems with the highest Mg/Fe values, the model favours accretion in the declining phase. In 13 systems is there a more than $3\sigma$ requirement for declining phase versus steady-state accretion, whilst in 54 systems is there a greater than $1\sigma$ requirement.

Those systems with moderate Mg/Fe are most likely to be in steady-state, whilst those systems where Mg/Fe is consistent with Mg/Fe seen in nearby stars, are found to be most likely to be in the build-up phase. The model, however, struggles to distinguish between build-up and steady-state, as the error bars are too large and encompass both possibilities. For no system is there a $>3\sigma$ requirement for steady-state over build-up (or vice versa).

\subsection{Evidence for Core-Mantle Differentiation}
\label{sec:coremantle}
 For most systems, Ca/Fe and Mg/Fe, can be adjusted by sinking effects, such that they are consistent within the error bars with the range of Ca/Fe and Mg/Fe seen in nearby stars. Thus, for the majority of systems (03832/202), the best model is the accretion of a primitive planetary body with no evidence for any alterations to its abundances, other than due to sinking effects.

\begin{figure}
    \centering
    \includegraphics[width=\columnwidth]{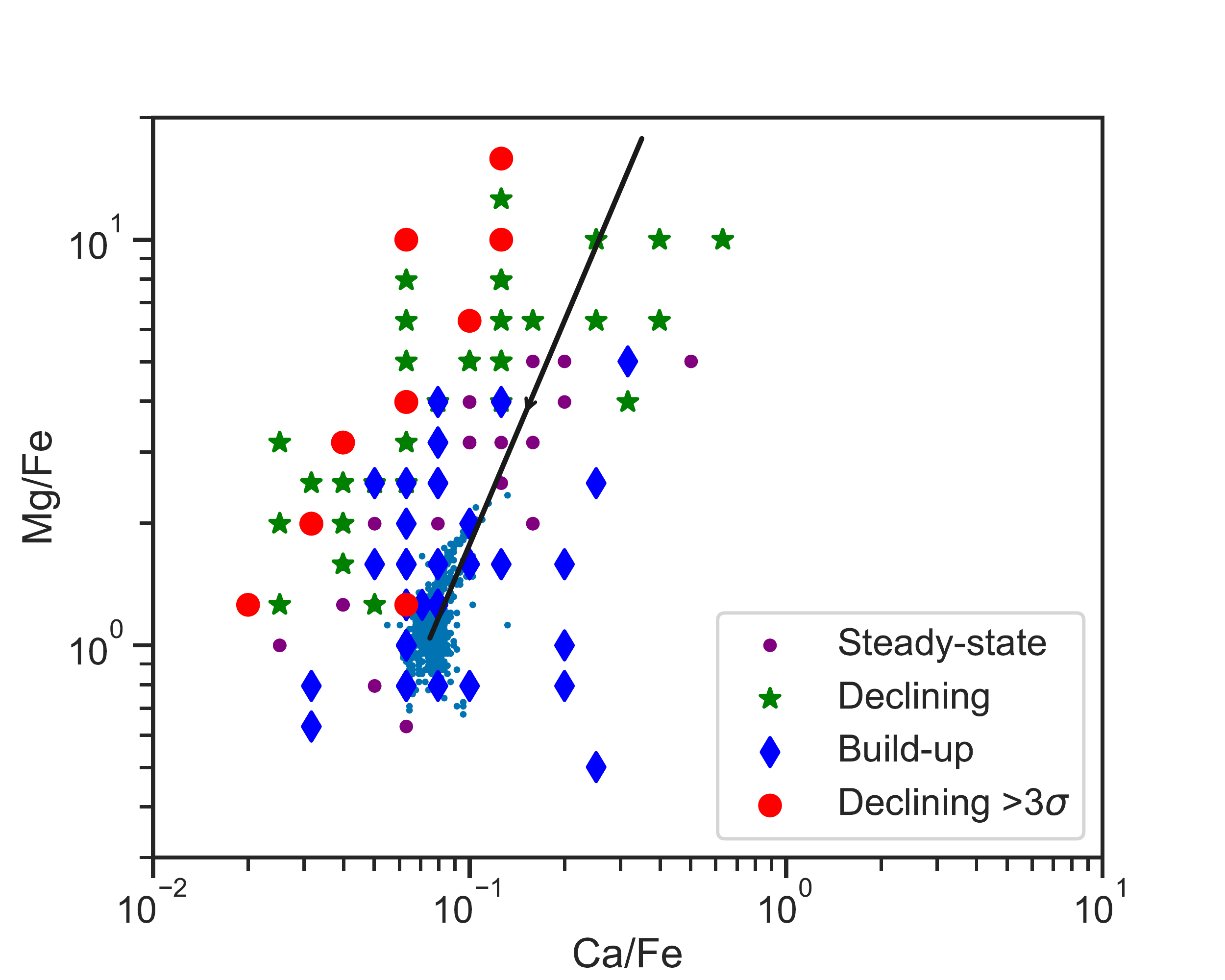}
    \includegraphics[width=\columnwidth]{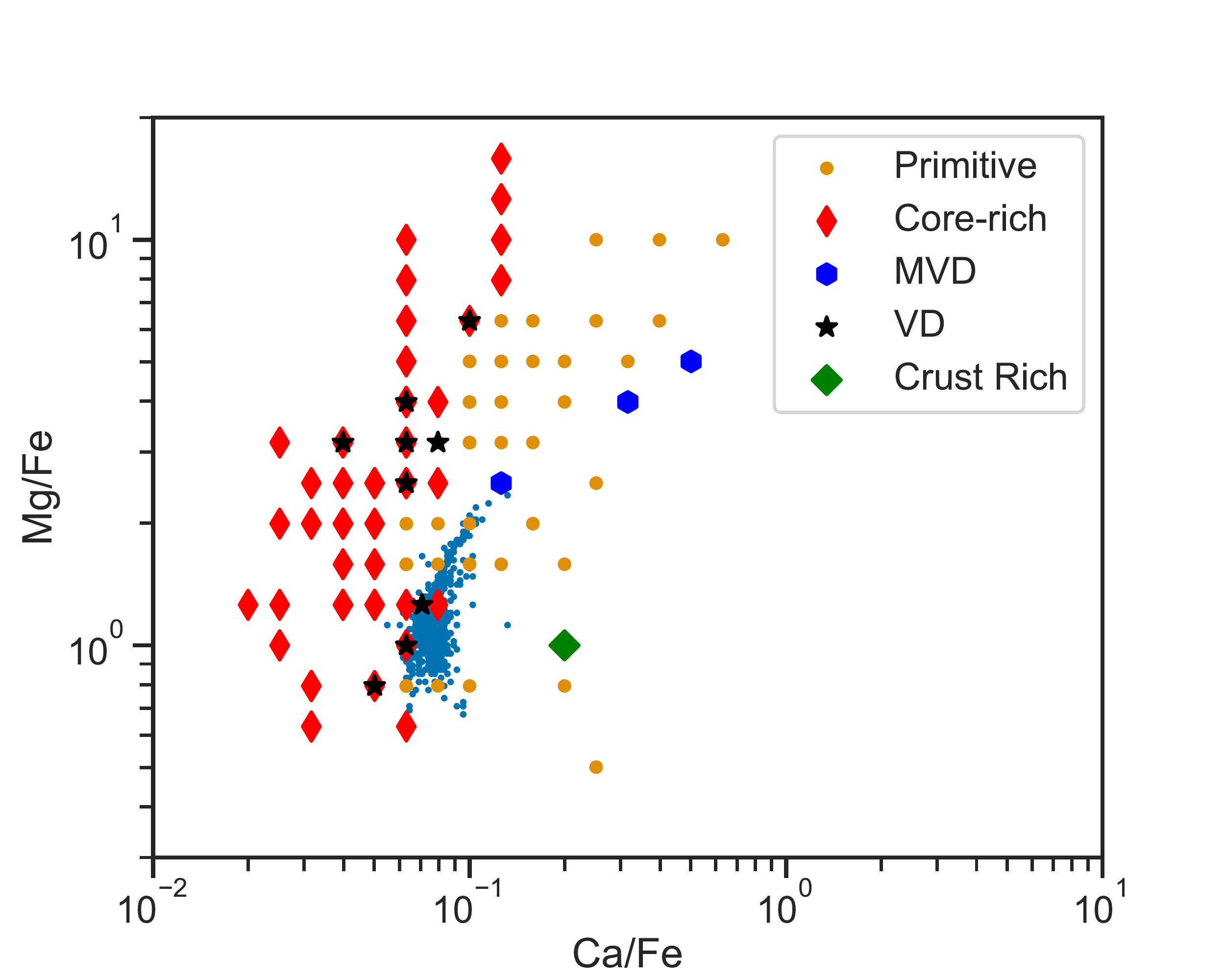}
        \caption{The Mg/Fe and Ca/Fe ratios in the abundances observed for all 202 systems. The stars in \citet{FischerBrewer2016} are shown in pale blue. The top panel shows those systems where the best model assumes accretion in the declining (green stars and red circles), steady-state (purple circles) or build-up phase (blue diamonds). Noting that whilst the declining phase is required to explain the abundances of 13 systems to $>3\sigma$, there is insufficient information to distinguish well between steady-state and build up phase. The black vector indicates the vector along which sinking alters abundances. The lower panel shows those models where the best-fit is Core-Rich (red diamonds), Crust-Rich (green diamond), VD (Volatile Depleted, black stars) or MVD (Moderate Volatile Depleted blue hexagons). There is insufficient information to find Mantle-Rich systems (see \S\ref{sec:coremantle} for a more detailed discussion.) Error ellipses are not included as they render the figure unreadable, however, typical values are on the order of 0.8 times the observed Mg/Fe or Ca/Fe.    }
    \label{fig:mg_fe_ca_fe}
\end{figure}

For those objects which have accreted iron-rich material, the accretion of a fragment of a larger differentiated planetary body, dominated by core material is often invoked as an explanation \citep{JuraYoung2014, Hollands2018}. Fig.~\ref{fig:mg_fe_ca_fe} shows that there are a handful of objects where the abundances sit below those of the \cite{FischerBrewer2016} stars in both Mg/Fe and Ca/Fe. The Bayesian model finds that these objects most likely accreted a core-rich body. Following a similar logic, planetary bodies with high Mg/Fe and Ca/Fe would be classified as mantle-rich. However, a mantle-rich body accreting in build-up is impossible to distinguish between a primitive body accreting in either in steady-state or the declining phase. The Bayesian model always favours sinking as an explanation, as this requires fewer free parameters in the model.

\begin{figure}
    \centering
    \includegraphics[width=1.15\columnwidth]{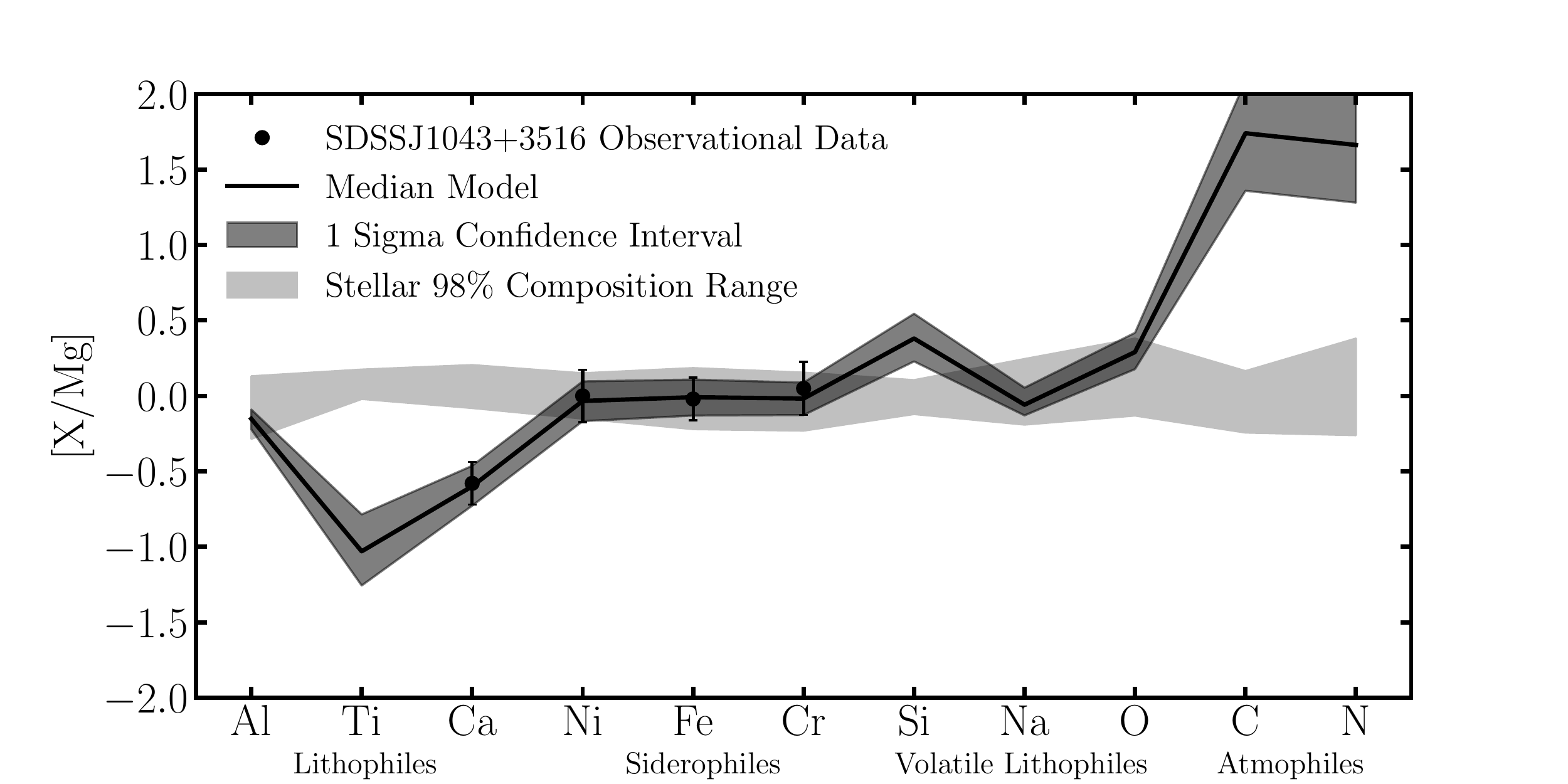}
    \caption{Core-rich body: In the same manner as Fig.~\ref{fig:nadepleted}, elemental ratios relative to Mg for the planetary material accreted by SDSSJ1043+3516, where Ca is depleted relative to Mg, such that the best explanation is the accretion of a core-rich body in the declining phase. The abundances of Cr and Ni corroborate this explanation. }
    \label{fig:corerich_ni}
\end{figure}

Those objects with an observed Ca/Mg below that of nearby stars cannot be explained by the accretion of material with stellar composition altered by sinking alone. In our simplistic model, the most likely explanation for these objects is that they have accreted a core-rich body in the declining phase. A detection of Cr or Ni can help support or refute this conclusion, as is shown by the example in Fig.~\ref{fig:corerich_ni} for SDSSJ1043+3516. For the majority of systems where only Mg, Ca and Fe are detected, there is insufficient information to categorically provide evidence for core-mantle differentiation.

\subsection{Evidence for Crustal Differentiation}

\begin{figure}
    \centering
    \includegraphics[width=1.15\columnwidth]{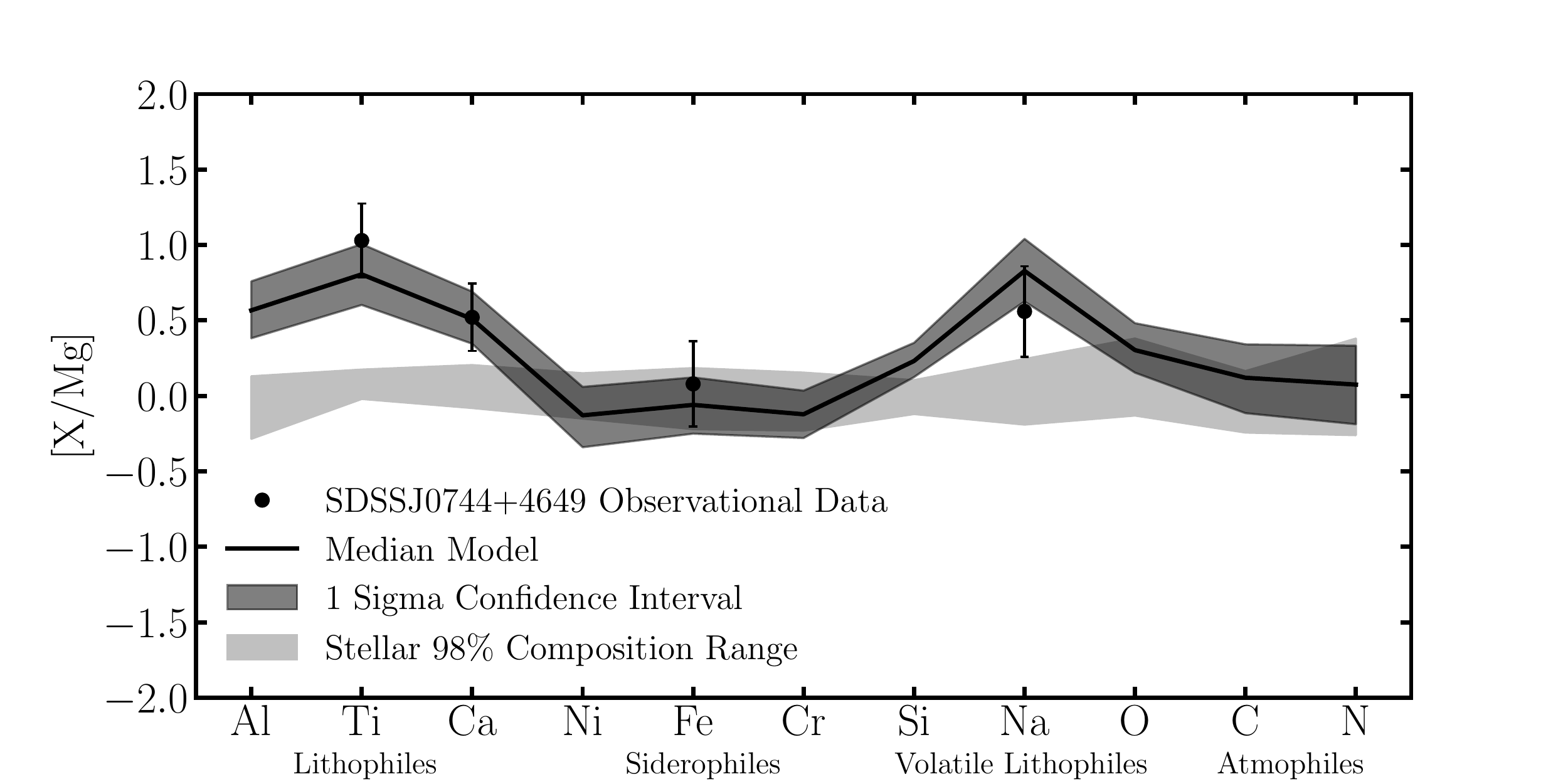}
    \caption{In the same manner as Fig.~\ref{fig:nadepleted}, elemental ratios relative to Mg for SDSSJ0744+4649, the high Na/Mg abundance in this system, as seen on Fig.~\ref{fig:na}, indicates that this system has likely accreted a crust-rich fragment. }
    \label{fig:crustrich}
\end{figure}

Crustal material is rich in Ca and Na. For those systems with only Ca, Fe and Mg detections, crustal differentiation cannot be investigated. This is because for Ca-rich systems, the effect of crustal differentiation cannot be distinguished from sinking or heating effects, both of which require fewer model parameters and are therefore more likely models. We therefore consider Na enrichment to be a better indicator of crustal differentiation. The only system which is rich in both Na and Ca is SDSSJ0744+4649, previously identified in \citep{Hollands2018} as crust-rich, shown in Fig.~\ref{fig:crustrich}. Our best-fit model suggests the accretion of a fragment consisting of 50\% crustal material. Two other systems that stand out as Na-rich can both be explained as being in the declining phase because Ca/Fe is not similarly enriched (SDSSJ0252+0054, SDSSJ2123+0016).

\begin{figure}
	\includegraphics[width=\columnwidth]{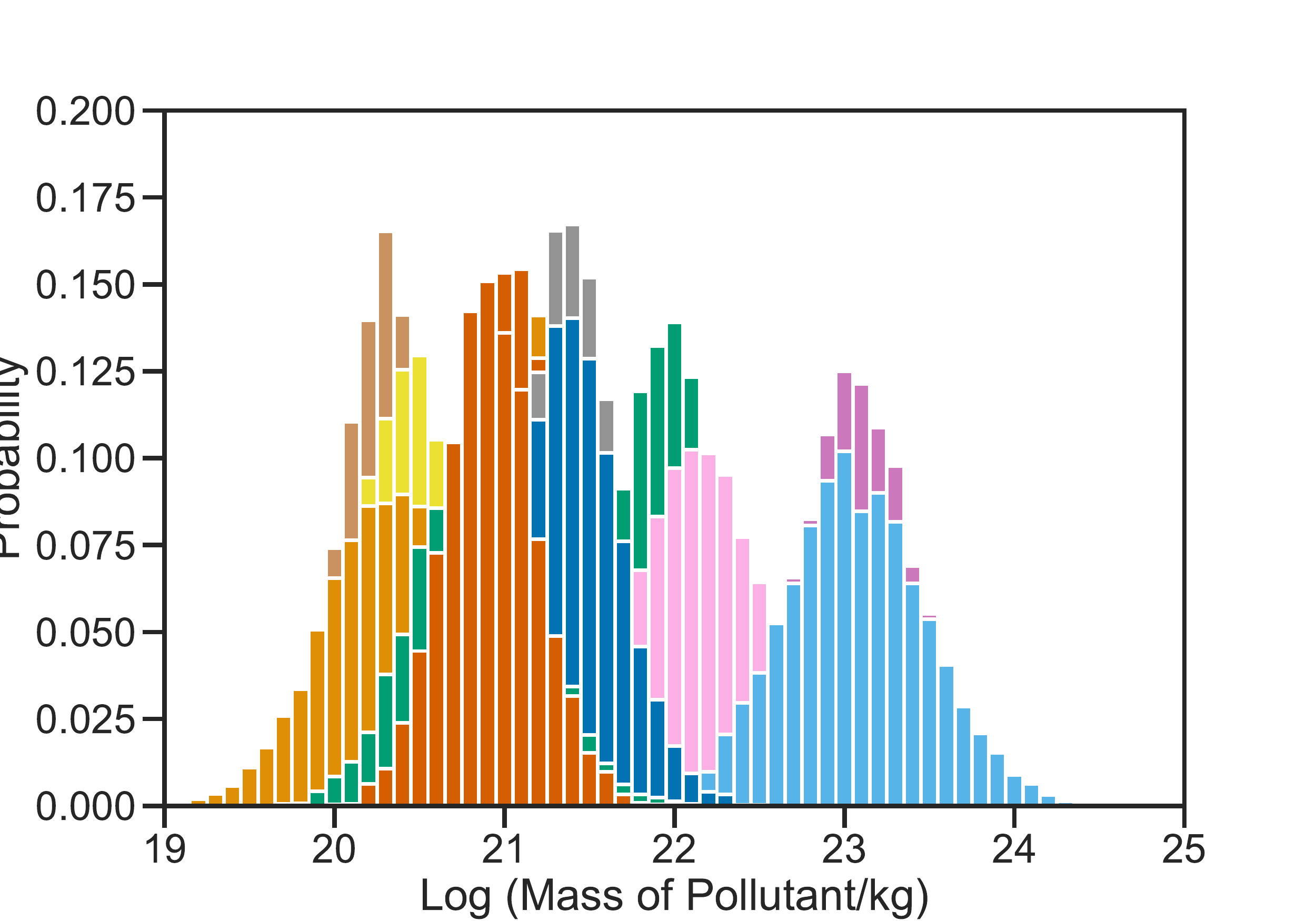}
 \caption{The posterior distributions for the total mass accreted onto the thirteen polluted white dwarfs whose abundances suggest that they are in the declining phase. During the declining phase, accretion has ceased and therefore, the mass currently in the atmosphere, alongside the sinking timescales, can be used to calculate the total mass accreted. The masses range from Vesta to the Moon.}
    \label{fig:decpost}
\end{figure}

\subsection{Constraints on the population}

Constraints on the population as a whole are useful for probing the origin of the `average' white dwarf pollutant. Such constraints can either be found by examining the frequency of each different best fit model or by summing the posterior distributions for each parameter for individual systems. It should be noted that both methods are subject to the observational biases in the sample, namely that high relative Ca and Fe abundances potentially generate larger colour shifts, thus, moderate volatile depleted, crust-rich, and core-rich pollutants may be more easily detected than primitive pollutants.

\begin{figure}
          \includegraphics[width=\columnwidth]{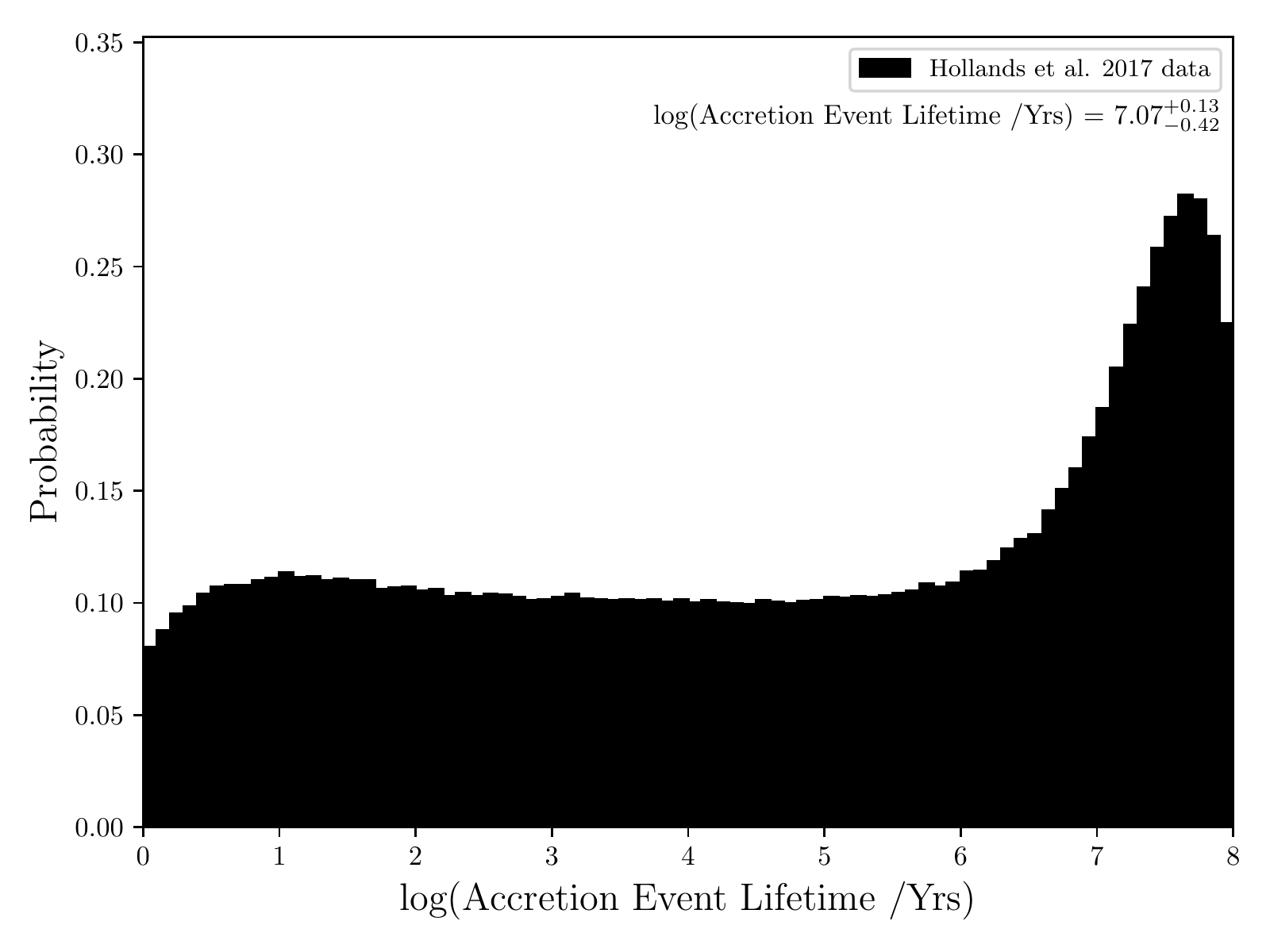}
     \caption{ The summed posterior distribution for the accretion event lifetime of all 202 systems.}
     \label{fig:lifetime}
\end{figure}

Table~\ref{fig:grid} summarises the most likely model fits. The top row counts the systems which have accreted geologically processed material, grouped by material type. The second row groups all 202 systems by the extent of formational heating, while the third row groups them by the phase of accretion.
Figure \ref{fig:lifetime} displays the summed posterior for the accretion event lifetime for all 202 systems, indicating that most systems are best explained by accretion that continued on timescales on the order of tens of Myrs.

\subsection{Results Summary}

The key results presented in this paper are:

\begin{itemize}[leftmargin=*]
\setlength\itemsep{1em}
\item The abundances observed in the majority (132/202) of the systems in the \cite{Hollands2017} sample are consistent with the accretion of pristine material, with the same abundances as the cloud of gas out of which the planetary bodies formed. 
\item (12/202) show evidence that the planetary bodies experienced heating, which depleted Na relative to more refractory species such as Mg and Ca
\item 3/202 systems show evidence that sufficient heating was experienced that moderately volatile species such as Mg, Fe, Na are depleted relative to refractory species such as Ca and Ti. 
\item In 54/202 systems the best model suggests that planetary material is accreted in the declining phase, i.e. accretion has finished, whilst for 13/202 there is a $>3\sigma$ for declining phase accretion over steady-state.
\item Such systems allow the masses of polluting bodies to be constrained, as it is known that no material is still to be accreted and resides in a reservoir close to the star. The masses found in this work range from bodies as massive as the Moon down to half that of Vesta. 
\item The constraints generated for the population as a whole on the accretion event lifetimes suggest that white dwarf accretion events last on average $11.7^{+4.1}_{-6.2}$\,Myrs. 
\item The best model for the abundances observed in 64 systems invokes geo-chemical differentiation followed by fragmentation. To a statistical significance of greater than 3$\sigma$, 8 systems require the accretion of a core-rich fragment, one systems require the accretion of a crust-rich fragment. Mantle-rich fragments are indistinguishable from sinking effects with only Ca, Mg and Fe detected.

\end{itemize}

\begin{table}
\centering
    \begin{tabular}{c|c|c|c|c}
    \hline
Core-rich	& Mantle-rich	 &Crust-rich \\
 63 ({\color{red} 8}) & 0 & 1 ({\color{red} 1})  \\\hline

Volatile Rich &	Volatile Depleted&	Moderate Volatile Depleted	\\
 190 {\color{blue} 38} &9  ({\color{red}  9})& 3 ({\color{red} 3}) \\ 
\hline
 Build Up Phase &	Steady State Phase	& Declining Phase	\\	
66	& 82({\color{red} 0})	& 54 ({\color{red} 13})\\ \hline

    \end{tabular}
    \caption{ The most likely model fits for the 202 systems considered here, alongside in red in brackets those systems where there is a more than $3\sigma$ requirement for the inclusion of the extra parameters. For the volatile depletion in blue are those systems where Na is detected. We note here that the current model cannot find mantle-rich fragments, as discussed in \S\ref{sec:coremantle}. }
    \label{fig:grid}
\end{table}

\section{Discussion}\label{diss}

The aim of this work is to improve the understanding of rocky exoplanetary bodies by constraining the origin of the planetary bodies which pollute the large sample of white dwarfs whose atmospheric abundances were derived in \cite{Hollands2017}. In order to constrain the origin of the planetary bodies, we use an upgraded version of the model presented in \cite{Harrison2018}. In this section we discuss the notable caveats of the model and the data and how they potentially affect the conclusions of this study.

\subsection{Discussion of caveats} \label{disscav}

The analysis presented here is based on the elements detected in \cite{Hollands2017}. For many objects there is relatively little information available as only Mg, Fe and Ca are detected. For such systems unless Ca is enhanced relative to Mg it is not possible to probe heating effects. Nor is it possible to distinguish mantle-rich fragments from sinking effects. Therefore, we conclude that most of these systems are primitive material which have not been heated substantially during formation, when the detection of further elements might verify a different reality. We note here that the model constrains the parameter values and the necessity of parameters to be included based on the observed abundances and the priors assumed.

In terms of the best model for the abundances of the accreted planetary bodies, the most limiting factor is our knowledge of how quickly different elements sink in the white dwarf atmosphere. This is crucial to the interpretation of the observed abundances, in particular the relative sinking of various elements. The physical conditions in the white dwarf envelope are the same for all elements, thus, in terms of relative sinking, their effects cancel out, and it is the average charges that are key to calculating the relative sinking of any pair of elements (as well as the ratio of their atomic weights). The new
calculations of \cite{Koester2020} include electron degeneracy,
non-ideal effects in the diffusion equation after \cite{Beznogov2013} and new calculations for the
collision integrals by \cite{Stanton2016}. The average charges for the elements are calculated with the
Saha equation (including non-ideal effects, the "lowering of the
ionization potentials") for low densities, and a Thomas-Fermi model
\citep{Feynman1949, More1985, Stanton2016} at
higher densities. This is a significant improvement over the simpler
estimates in \cite{Paquette1986} adapted from the
original work of \cite{Fontaine1979}. The new
average charges for Si and Ca in He compare favorably with the much
more sophisticated calculation for these special cases by \cite{Heinonen2020}.

The model does not include any errors on the sinking timescales, as they are difficult to quantify, however, we note that the difference in sinking time ratios depending on the convective overshoot is small compared to differences based on the ionisation prescription. For example, the sinking time ratio for Mg/Fe decreases from 2.71 using the \citep{Paquette1986} model to 2.20 (without overshoot) or 2.24 with one pressure scale height overshoot in the \cite{Koester2020} models used here (for a DZ of 6,000K and log(g)=8). In particular, we note that the conclusion that some fragments are best explained by the accretion of a core-rich fragment in the declining phase is most influenced by the exact ratio of sinking timescales. We can, however, be confident that sinking alone cannot explain the full spread in observed abundances seen here. The choice of one pressure scale height does, nonetheless, strongly influence the calculated accreted masses, which are of less relevance here.

Crucially, we note here that the estimation of the uncertainties on the observed abundances is key to our Bayesian model conclusions. Given that the uncertainities on these, alongside the sinking timescales, are currently difficult to quantify, the results of the models presented here may change in the future, based on changes to these crucial input data. In the current work, we primarily present a powerful new method that can in the future easily
be adapted to new or improved observations.

\subsection{Validity of the model} \label{dissval}

The major caveat regarding the modelling of planetesimal abundances is the assumption that a planetesimal composition is dictated when it condenses out of a protoplanetary disc in chemical equilibrium. While this is a simplistic assumption, the model can reproduce the major element composition of the rocky bodies in the Solar System \citep{Moriarty2014, Harrison2018}. Other processes could remove volatiles and create temperature related signatures on planetary bodies. Heating on the giant branches has shown to be ineffective for large asteroids \citep{jura2010, Malamud2016}, however, collisionally induced post-nebula volatilisation processes could affect the abundances of the planetesimals modelled \citep{Palme2003, Oneill2008}. Post-nebula volatilisation is not taken into account in our model, however, given the elements utilised in this study as such volatilisation effects will change the abundances observed in a similar manner to nebula condensation. Therefore, we note that the temperatures constrained should be thought of as the maximum temperatures experienced by the body rather than strict formation temperatures. 

The disc model used assumes no vertical mixing and uses an alpha parameterisation. The use of this simplistic model is not expected to dramatically effect the results as it is only used to convert protoplanetary disc formation times and distances into temperatures and pressures. Therefore, even if the exact conversion is not robust the temperatures constrained should remain reliable. 

Planetesimal formation is far more complex than our model suggests and planetary migration and dust migration are potentially crucial effects that are not included in our model. However, given the model's ability to reproduce the abundances of the Solar System bodies and the majority of the polluted white dwarfs studied thus far, the inclusion of a more sophisticated disc, chemistry, and planetesimal formation model will likely not affect the results and is likely not necessary.

In this work as in \cite{Harrison2018} we consider an Earth-centric differentiation model which assumes that the cores of planetary bodies form with the same composition as the Earth's core. We show that such a model can reproduce the abundances of the majority of white dwarf pollutants, given the current observational errors. We do, however, note that core-mantle differentiation may occur under different conditions in exoplanetesimals. 
We note here that we do not consider the possibility of thermohaline instabilities when converting planetesimal abundances to the atmospheric abundances. However, such instabilities mostly influence hotter white dwarfs than those considered here \citep{Deal_2013, Wachlin2017, Bauer2018}.

The main caveats involving the statistical framework used in this work involve the number of live points selected and the choice of prior distributions. The number of live points chosen for each model run was 2,000 as this was found to be sufficient to calculate the log Bayesian evidence to an uncertainty of the order 0.1. Increasing the number of live points would no doubt decrease this uncertainty, however, it is already small enough that any changes would not affect the results presented. The majority of the priors chosen were fixed by the model, however, feeding zone size, time since accretion started, accretion event lifetime, and pollution fraction were not. The feeding zone size, time since accretion started, and pollution fraction were tested for various ranges and it was found for the white dwarf sample analysed in this work the ranges were sufficient for all the systems which required them to converge as well as being physically realistic. The range of accretion event lifetimes used were based on previous studies \citep{Girven2012, Wyatt2014, Ribas2015}, however, changes to these ranges would have a small effect on the results, relative to the other parameters considered.

\subsection{Discussion of results} \label{dissres}

In this work the majority of systems in the \cite{Hollands2017} sample are found to be easily reproduced by a scenario in which the star is accreting primitive unprocessed and unheated material. It is unsurprising that the accreted material is often primitive, and thus, has refractory abundances which match those expected for its original planetesimal forming disc, as the majority of Solar System bodies have refractory abundances which match that of the solar photosphere, and thus their natal disc. 

\subsubsection{Constraints on formation temperature}

Our study provides evidence for the occurrence of heating and the depletion of volatiles in the planetary material accreting onto many white dwarfs. Na is the crucial element. Na has a condensation temperature of $\sim$1,000\,K. Therefore, Na abundances similar to that observed in stellar photospheres are indicative of formation temperatures lower than $\sim$1,000\,K, while depleted abundances can constrain formation temperatures higher than $\sim$1,000\,K. 38 systems have Na abundances consistent with stellar, while 12 systems have depleted Na abundances. Such systems are potential probes of the formation temperature of white dwarf pollutants and thus, the formation location of exo-planetary bodies in white dwarf planetary systems. If the temperatures are converted into formation locations in a protoplanetary disc around an A-type star (Section \ref{HDFSE}), 38 systems are required to have accreted bodies which formed exterior to $\sim$5\,AU. 12 systems are required to have accreted bodies which formed interior to $\sim$5\,AU. Therefore, as previously concluded in \cite{Harrison2018} white dwarf pollutants have a range of formation locations.

3 systems analysed in this work require not just incomplete condensation of Na but also incomplete condensation of the moderate-volatiles such as Fe and Mg. Systems whose pollutants have experienced such extreme heating have been discussed before \citep{Harrison2018} and are especially interesting as they have no Solar System equivalent \citep{Xu2013}. If the formation temperatures constrained are converted into formation locations in a protoplanetary disc around an A-type star, these bodies are required to form interior to 2\,AU. As the inner regions of post-main sequence systems are cleared of large asteroids during the giant branches, due to stellar engulfment or due to the asteroids being spun up to break-up by the YORP effect \citep{VerasYORP, Veras2016}, potentially such bodies are not expected to survive to the white dwarf phase. A quick estimate using the expected mass of the pollutant bodies validates that this material could survive to the white dwarf phase, if it formed earlier than $\sim$0.12\,Myrs after the formation of the star's protoplanetary disc.

\subsubsection{Constraints on differentiation}

The accretion of fragments of differentiated bodies onto white dwarfs has been proposed due to atmospheric abundances which are rich in the siderophiles/lithophiles \citep{Zuckerman2011, Xu2013, Wilson2015}. The Bayesian model presented here is able to find statistically significant evidence ($>3\sigma$) for differentiation in 9 systems, of which 8 have accreted core-rich fragments and one a crust-rich fragment. As discussed in \S\ref{sec:coremantle}, the combination of sinking timescales is such that with only Ca, Mg and Fe detected, it is not possible to distinguish mantle-rich fragments from primitive bodies accreting in the declining phase. Whilst the model will always consider the declining phase as a better model, as it has fewer free parameters, given the 8 core-rich fragments detected, it remains plausible that 8 mantle-rich fragments are hidden amongst the 22 objects in the declining phase with Ca/Fe $>$ Ca/Fe$_\odot$. This would fit with many collision models \citep{Carter2017} which predict equal quantities of core and mantle fragments. On the other hand, uncertainties on the ratios for sinking timescales of different elements may be sufficient to rule out core-mantle differentiation as an explanation for any systems. 


We note here that whilst Ni and Cr can corroborate or refute a core-rich model, the Bayesian model presented here does not correctly predict the Ni and Cr abundances in the core of an asteroid-sized object, where core-mantle differentiation will have occurred at significantly lower pressures than in the centre of our own planet, whose abundances are currently used.

The strength of our analysis is that we can provide statistical significances for the requirement of each system to have accreted fragments of differentiated bodies. \cite{Hollands2018} identified 4, 3, and 2 systems as having extreme Ca, Fe, and Mg abundances and attributed this to the accretion of crust-rich, core-rich, and mantle-rich fragments. Our analysis finds that of the Ca-rich systems outlined by \cite{Hollands2018} only 1 can be attributed to being crust-rich (SDSSJ0744+4649 to 5$\sigma$ significance), whilst for 2 other systems (SDSSJ1033+1809 and SDSSJ1351+2645) there is insufficient information available for a crustal model to be the best fit. SDSSJ1055+3725 was no longer included in the sample (see \S\ref{PWDD}).
The 3 Fe-rich systems (SDSS1043+3516 ($5.2\sigma$), SDSSJ0741+3146 ($2.8\sigma$) and SDSSJ0823+0546 ($10\sigma$) are indeed found to be core-rich. Both the two 2 Mg-rich systems, SDSSJ0956+5912 and SDSSJ1158+1845 actually have fairly typical Mg/Fe ratios and due to the combination of sinking timescales, the best-fit models suggest that both systems are core-rich to $2\sigma$.

\subsubsection{The mass of white dwarf pollutants}

For those systems in the declining phase, accretion has finished and the mass in the atmosphere, combined with the sinking timescales can be converted into the total mass of the polluting body. This calculation no longer yields a lower limit as there cannot be any material left in a circumstellar orbit which is yet to accrete. Such systems are a unique probe into the mass of the polluting bodies and, as shown in Figure \ref{fig:decpost}.   The bodies polluting white dwarfs are found to have masses between just half that of Vesta and the mass of the Moon. The bodies polluting white dwarfs are likely to be similar to large Solar System asteroids, as previously concluded in the literature \citep{Farihi2010, Girven2012}, although at least 2 systems have accreted planetary bodies approaching the mass of the Moon. 

\subsubsection{Constraints on the accretion event lifetimes}
The accretion event lifetime for a system can be constrained from the abundances because the duration of accretion affects the probability that the system will be observed in build-up, steady state, or declining phase. The accretion event lifetimes from the 202 systems analysed here suggests that the average accretion events last $log(t_{\rm event}) = 7.07^{+0.13}_{-0.42}$ \,yrs. This result is consistent with the result derived in \cite{Girven2012} and is consistent with the theoretical predictions of \cite{rafikov2} and \cite{Rafikov1}, but is in stark contrast to the short accretion event (or disc) lifetimes predicted by \cite{Wyatt2014} to explain the difference between observed accretion rates of DA and DB white dwarfs.

\section{Conclusions}

In this work we present a Bayesian model to explain the abundances of planetary material observed in the atmospheres of some white dwarfs. Our method reproduces the abundance patterns observed in polluted white dwarf atmospheres by modelling the chemical composition expected if the white dwarf stars accreted planetesimals which could form in protoplanetary discs with a range of initial compositions, at various temperatures, and with various geological and collisional histories. 

The relative sinking of the observed elements in the white dwarf atmosphere influences their observed  abundances. The model finds the most likely original abundances of the accreted material. As such, the model is able to provide some key insights regarding the interpretation of the observed abundances, however, we note here that these results depend strongly on the input data. Given that many of the errors are currently difficult to quantify exactly, the main purpose of this work is to present a new powerful method that can readily be applied to new or improved observations in the future. 

Our model is able to explain all 202 systems in the \cite{Hollands2017} sample to $\chi^{2}$ per element values of less than one. We use this as evidence that the initial composition of the planetesimal forming disc, formational heating, and differentiation and collisions are the key processes that determine the abundances of exo-planetary bodies, as is the case in our Solar System. 

In our Solar System, the refractory abundances of most meteorites are consistent with solar abundances. In this work, we provide observational evidence that many exo-planetesimals have refractory abundances that match those of their host-stars. This is shown by the 132/202 systems whose abundances fit within the range of abundances seen in nearby stars. A match between stellar and planetary abundances is crucial in determining the composition of exoplanets \citep[e.g.][]{Dorn2017}. Our findings validates this hypothesis.

We show that exoplanetary bodies experience similar thermal processing to the planets, meteorites, and Ca-Al inclusions in our Solar System. The composition of the material that pollutes 11 white dwarfs systems is required to have experienced significant heating during formation in order to explain the atmospheric abundances. 8 systems require such heating to have caused the incomplete condensation of the volatile species, such as Na, while 3  systems require such heating to have caused the incomplete condensation of the moderate-volatile species, such as Mg.

Our analysis supports the idea that differentiation and collisions are common processes in exo-planetary systems and that differentiation occurs in a similar manner in exo-systems as it does in the Solar System. The abundances reported for 65/202 systems suggest the accretion of fragments of core-mantle differentiated bodies, with 8 systems requiring a core-rich body to $>3\sigma$ and one system the accretion of a crust-rich body. 

We find that white dwarf pollutants are typically the size of large asteroids. The abundances for thirteen systems require accretion in the declining phase to $>3\sigma$, thus the total mass accreted during the accretion event can be determined. These masses range from bodies as massive as the Moon down to half that of Vesta. 

The accretion event lifetimes for the population of pollutants can also be constrained from their abundances (Figure \ref{fig:sink}) and in this work the accretion events were constrained to last $11.7^{+4.1}_{-6.2}$\,Myrs on average. Such a result is consistent with the observational properties of polluted white dwarfs \citep{Girven2012} and theoretical predictions of white dwarf pollution \citep{rafikov2,Rafikov1}.

\section{Data Availability}
The codes used in this work will be made available upon reasonable request to the authors. The results of the models are presented in the Supplemenetary Information.

\section*{Acknowledgements}
We would like to thank the referee for helpful comments that improved the quality of this manuscript. 
We would like to thank Mark Hollands for his helpful, useful, and insightful comments regarding the white dwarf data set and provision of the WHT spectra for a handful of objects. We would also like to thank Ryan MacDonald for insightful discussions regarding the implementation of the nested sampling algorithm and Oliver Shorttle for his helpful discussion regarding the differentiation model. JHDH is grateful to the Science \& Technology Facilities Council for a PhD stipend. AB acknowledges the support of a Royal Society Dorothy Hodgkin Fellowship. AMB and AB acknowledge the support of a Royal Society Enhancement Award, RG92907. S.B. acknowledges support from the Laboratory Directed Research and Development program of Los Alamos National Laboratory under project number 20190624PRD2. M.K. gratefully acknowledges support from the EU under the Marie Sklodowska-Curie Fellowship grant agreement No. 753799 and the University of Tartu ASTRA project 2014-2020.4.01.16-0029 KOMEET.



\bibliographystyle{mnras}
\bibliography{JHref}



\bsp	
\label{lastpage}
\end{document}


\appendix
\newcommand{\hbAppendixPrefix}{A}
\renewcommand{\thefigure}{\hbAppendixPrefix\arabic{figure}}
\setcounter{figure}{0}
\renewcommand{\thetable}{\hbAppendixPrefix\arabic{table}} 
\setcounter{table}{0}
\renewcommand{\theequation}{\hbAppendixPrefix\arabic{equation}} 
\setcounter{equation}{0}

\onecolumn
\begin{centering}
\begin{longtable}{l cccc cccc cccc cccc}
\caption{\label{tab:abundances} Abundances of all elements used in this work from \citet{Blouin2020} and scaled from \citet{Hollands2017}, as described in \S2.1 Abundances relative to helium are given in dex, with $1\sigma$ errors in the adjacent column. $^*$ Ti, Ni, Cr or Na detection not included, as the detections were not clear during the re-analysis. $^\dagger$ Cr, Ti, Na or Ni abundances from a re-analysis, rather than scaling.  } \\
\hline
         SDSSJ &   $T_{\rm eff}$[K] &  log(g) & Ti & &  Ca &  & Ni &  &  Fe & & Cr &   &  Mg &    & Na &   \\ \hline

\hline
\endfirsthead
\hline

 SDSSJ &   $T_{\rm eff}$[K] &  log(g) & Ti & &  Ca &  & Ni &  &  Fe & & Cr &   &  Mg &    & Na &   \\ \hline

\hline
\endhead
\hline
\endfoot
 0002+3209 &  6466 &    8.26 &        - &            - &      -9.10 &             0.20 &        - &            - &      -7.90 &             0.20 &        - &            - &      -7.40 &             0.20 &        - &            - \\
 0004+0819 &  5843 &    8.00 &        - &            - &      -8.80 &             0.20 &        - &            - &      -7.60 &             0.40 &        - &            - &      -7.20 &             0.20 &        * &            * \\
 0006+0520 &  5783 &    7.84 &        - &            - &      -9.80 &             0.20 &        - &            - &      -9.60 &             0.30 &        - &            - &      -8.60 &             0.50 &        - &            - \\
 0010-0430 &  6903 &    8.12 &        - &            - &      -8.50 &             0.10 &        - &            - &      -7.00 &             0.10 &        - &            - &      -6.70 &             0.10 &        - &            - \\
 0013+1109 &  6090 &    8.00 &        - &            - &      -9.20 &             0.40 &        - &            - &      -8.00 &             0.50 &        - &            - &      -7.50 &             0.40 &        - &            - \\
 0019+2209 &  5797 &    7.61 &        - &            - &      -9.70 &             0.20 &        - &            - &      -8.50 &             0.20 &        - &            - &      -8.40 &             0.20 &     -9.78 &             0.15 \\
 0044+0418 &  6104 &    8.22 &        - &            - &      -9.80 &             0.10 &        - &            - &      -8.40 &             0.20 &        - &            - &      -8.30 &             0.30 &        - &            - \\
 0046+2717 &  8053 &    8.46 &        - &            - &      -7.20 &             0.20 &        - &            - &      -6.30 &             0.30 &        * &            * &      -5.40 &             0.20 &        - &            - \\
\end{longtable}
\end{centering}
\newpage

\onecolumn

\begin{centering}

\begin{longtable}{ l c c c c c c c c c c c  }

\caption{The model with the highest evidence for each system. P: primitive, Core: core-rich fragment, Mantle: mantle-rich fragment, Crust: crust-rich fragment, VR: no evidence for volatile depleted, VD: volatiles (Na) depleted, Hot: moderate volatiles depleted, BU:build-up phase, SS:steady-state phase, Declining: Declining phase } \label{tab:best-fit} \\

\hline   

\hline   


Name & Model & P &  Core&  Mantle  &  Crust & VR & VD & Hot &  BU &  SS  &  Declining \\ \hline

\endfirsthead

\multicolumn{3}{c}%
{{\bfseries \tablename\ \thetable{} -- continued from previous page}} \\
\hline  

Name & Model & P &  Core&  Mantle  &  Crust & VR & VD & Hot &  BU &  SS  &  Declining \\ \hline

\endhead

\endhead
\hline
\endfoot

 0002+3209 &    M1 &        Y &       N &         N &        N &            Y &               N &                        N &            N &                 Y &             N \\
 0004+0819 &    M4 &        Y &       N &         N &        N &            Y &               N &                        N &            N &                 Y &             N \\
 0006+0520 &    M1 &        Y &       N &         N &        N &            Y &               N &                        N &            N &                N &              Y \\
 0010-0430 &    M4 &       N &        Y &         N &        N &            Y &               N &                        N &            N &                N &              Y \\
 0010-0430 &    M4 &       N &        Y &         N &        N &            Y &               N &                        N &            N &                N &              Y \\
 0010-0430 &    M4 &       N &        Y &         N &        N &            Y &               N &                        N &            N &                N &              Y \\
 0010-0430 &    M4 &       N &        Y &         N &        N &            Y &               N &                        N &            N &                N &              Y \\
 0010-0430 &    M4 &       N &        Y &         N &        N &            Y &               N &                        N &            N &                N &              Y \\
\end{longtable}
\end{centering}
\newpage

\onecolumn

\begin{centering}

\begin{longtable}{ lllllllllllllll }

\caption{The best fit model and model parameters for each system. The evidence is quoted for the best-fit model ($\mathcal{Z}_{\textrm{best}}$ ), alongside the median values of the parameters and a requirement for core-mantle differentiation ($\sigma_{\rm core}$), a requirement for steady-state versus build-up phase accretion ($\sigma_{\rm SS}$), a requirement for declining phase versus steady-state Accretion ($\sigma_{\rm dec}$) and a requirement for incomplete condensation ($\sigma_{\rm incomplete}$), quoted in $\sigma$. The model parameters are time since accretion started (t in Myrs), the distance at which the observed chemistry would be produced in a \citet{Chambers2009} disc in chemical equilibrium, $d_{\rm formation}$ in au, $z_{\rm formation}$ the width of the formation zone, in au, $t_{\rm event}$ the lifetime of the accretion event in log (yrs), $f_c$ the number fraction of core material, $f_o$ the number fraction of crustal material, Z the total sum of the observed pollutants in log (dex). \label{tab:param} }\\

\hline  Name & Model &  $\mathcal{Z}_{\textrm{best}}$  &  $\chi^2$ & $\sigma_{\rm core}$ & $\sigma_{\rm SS}$ & $\sigma_{\rm dec}$ & $\sigma_{\rm incomplete}$ &   t &  $d_{\rm formation}$ & $z_{\rm formation}$ &  $t_{\rm event}$  &  $f_c$ & $f_o$  &  Z\\ \hline 
\endfirsthead

\multicolumn{6}{c}%
{{\bfseries \tablename\ \thetable{} -- continued from previous page}} \\
\hline  Name & Model &  $\mathcal{Z}_{\textrm{best}}$  &  $\chi^2$ & $\sigma_{\rm core}$ & $\sigma_{\rm SS}$ & $\sigma_{\rm dec}$ & $\sigma_{\rm incomplete}$ &   t &  $d_{\rm formation}$ & $z_{\rm formation}$ &  $t_{\rm event}$  &  $f_c$ & $f_o$  &  Z \\ \hline

\endhead
\hline
\endfoot
 0002+3209 &    M1 &     -2.54 &         0.46 &       - &    1.60 &      - &                                     - &                       2.86 &                         - &                   - &                                6.88 &                     - &                      - &               -7.37 \\
 0004+0819 &    M4 &     -2.83 &         0.00 &       - &    1.40 &      - &                                     - &                       5.69 &                         - &                   - &                                4.38 &                    0.43 &                      - &               -7.01 \\
 0006+0520 &    M1 &     -3.34 &         0.38 &       - &     - &     2.18 &                                     - &                      27.68 &                         - &                   - &                                3.86 &                     - &                      - &               -8.13 \\
 0010-0430 &    M4 &     -2.68 &         0.00 &      3.68 &     - &     3.21 &                                     - &                       4.00 &                         - &                   - &                                3.24 &                    0.70 &                      - &               -6.52 \\
 0013+1109 &    M4 &     -3.42 &         0.00 &      1.12 &    1.20 &      - &                                     - &                       8.69 &                         - &                   - &                                4.24 &                    0.51 &                      - &               -7.30 \\
 0019+2209 &    M1 &     -2.36 &         0.05 &       - &     - &      - &                                     - &                      15.99 &                         - &                   - &                                4.34 &                     - &                      - &               -8.14 \\
\end{longtable}
\end{centering}

\bibliographystyle{mnras}
\bibliography{JHref}


\bsp	
\label{lastpage}